\documentclass{jfm}

\usepackage{graphicx}
\usepackage{newtxtext}
\usepackage{newtxmath}
\usepackage{natbib}
\usepackage{hyperref}
\hypersetup{
    colorlinks = true,
    urlcolor   = blue,
    citecolor  = black,
}

\newcommand{\RomanNumeralCaps}[1]
\linenumbers

\usepackage{xcolor}
\usepackage[draft,inline,nomargin,index]{fixme}

\fxsetup{theme=color,mode=multiuser}
\FXRegisterAuthor{sv}{asv}{\color{red}Me}
\FXRegisterAuthor{aa}{aaa}{\color{green}Other}

\title{Rayleigh-Taylor instability in binary fluids with miscibility gap}

\author{Anubhav Dubey\aff{1}
  \corresp{\email{anubhav.dubey@u-bordeaux.fr}},
  Constantin Habes\aff{2},
  Holger Marschall\aff{2}
 \and Sakir Amiroudine\aff{1}}

\affiliation{\aff{1}Univ. Bordeaux, CNRS, Bordeaux INP, I2M, UMR 5295, F-33400, Talence, France
\aff{2}Department of Mathematics, Computational Multiphase Flow, Technical University Darmstadt, Darmstadt, Germany}

\begin{document}
\maketitle

\begin{abstract}
A novel phase field method is proposed to model the continuous transition of binary fluids exhibiting temperature sensitive miscibility gap, from immiscible state to miscible state via partially miscible states. The model is employed to investigate the isothermal single-mode Rayleigh-Taylor (RT) instability for binary fluids as the system temperature is varied. Assuming potential flow and utilizing Boussinesq approximation, we derived the dispersion relation for gravity-capillary waves and the RT instability. The study reveals the early-stage growth characteristics of the interfacial perturbation. Three zones with distinct qualitative behaviour for the growth rate are identified as a function of Atwood number and Weber Number. Subsequently, Boussinesq approximation is relaxed to obtain coupled Cahn-Hilliard-Navier-Stokes equations to perform numerical simulations. The results from the numerical simulations corroborate the findings from the dispersion relation at early-stages. \\
Further investigation of the late-time dynamics for viscous fluid pair reveal the tortuous topology presumed by the interface. The emanation of secondary instability in form of Kelvin-Helmholtz rolls is observed. The formation of Kelvin-Helmholtz rolls is found to be dependent on the system temperature. Finally, we present the effect of the slow nature of diffusion process.

\end{abstract}


\section{Introduction}
\label{sec:Intro}

 Rayleigh-Taylor (RT) instability is the evolution of a perturbed interface between two superposed fluids subjected to an external acceleration directed from heavier fluid to lighter fluid. The RT instability is relevant for a wide range of natural and industrial processes such as flame acceleration in supernovae \citep{Bell2004}, weather inversion \citep{Chen2015}, formation of Mammatus clouds \citep{Ravichandran2020}, geophysical flows \citep{Mao2024}, magneto-hydrodynamic power generators \citep{Cole1973}, inertial confinement fusion \citep{Haan2011}, designing soft materials \citep{Marthelot2018} to enumerate a few. For fluids with unstable density stratification, \citet{Rayleigh1882} and \citet{Taylor1950} developed the linear theory to demonstrate the growth of all infinitesimal perturbations in absence of viscosity and surface tension. The theoretical results were corroborated experimentally by \citet{Lewis1950} for low viscosity fluids, during early stages of perturbation growth. The experiments revealed the formation of long thin spikes of the heavier fluid interspersed with rising bubbles of the lighter fluid. Nevertheless, the choice of large accelerations and relatively small wave numbers fell short of quantifying the effects of the surface tension. \citet{Bellman1954} incorporated the effects of surface tension and viscosity while analyzing the linear perturbation growth. Surface tension introduces a threshold wave number for the manifestation of the instability. An initial perturbation with a wavelength lower than the cutoff leads to oscillatory temporal evolution of the interface due to over-stability. This wavelength is independent of the viscosity \citep{Chandrasekhar1961} of the fluids. \citet{Cole1973} reported reasonable agreement between the linear theory and the experimentally obtained growth rate at water-air interface during early-time stage.
 
  The asymmetrical growth of the surface \citep{Emmons1960} in form of narrowing of the crest and broadening of the trough of the wave accentuated the need of large amplitude non-linear analysis. \cite{Menikoff1983} employed the potential theory aided with conformal mapping to reveal the velocity and acceleration of the spike and bubble at late-time stages. This temporal evolution of the interface morphology underpins the RT mixing. \cite{Youngs1984} numerically segregated the process of RT mixing into three successive stages based on the state of perturbation growth. These stages can be enumerated as the exponential growth regime, constant velocity growth and the quadratic time dependent growth. \cite{Read1984} corroborated the presence of quadratic time dependent growth experimentally. The study, furthermore, demonstrated the challenges associated with experimental investigation of the RT instability which entails dynamics spanning across various spatio-temporal scales. \citet{Tryggvason1988} proposed an Eulerian-Lagrangian vortex method to simulate RT instability between inviscid fluids under zero surface tension assumption. The study highlighted the dependence of interface evolution on the wavelength of the perturbation and the density ratio between the fluids. In non-ideal (purely-viscous) fluids, the Stokes analysis \citep{Newhouse1990} performed under the assumption of creeping flow, establishes viscosity ratio as one of the governing parameters for the non linear evolution dynamics. An increase in surface tension suppresses the effects of the viscosity ratio. \citet{Elgowainy1997} extended the analysis of finite liquid layers by relaxing the creeping flow assumption and solving the complete Navier-Stokes equations within a two-dimensional framework deploying the volume-of-fluid method \citep{Hirt1981}. The competition between the driving forces behind the growth of bubble and spike lead to asymmetry in the interface morphology. These forces primarily depend on the geometrical parameters (initial amplitude and wavelength of perturbation) and the fluid properties (viscosity and surface tension). 
 
 The dependence of the flow features on the geometrical parameters leads to scaling issues \citep{Cook2001} pertinent to RT instability at the interface between miscible fluids, as the interface evolution is subjected to relative competition between diffusive and convective time scales. The diffusivity between the miscible fluids renders the effective density ratio and the surface tension play a dynamic role thereby necessitating a complete direct numerical simulation incorporating species diffusion. \citet{Young2001} assumed the diffusive time scale for miscible fluids to be much larger to maintain a constant interface thickness. The study revealed the existence of an evolutionary stage marked by the slow fall of the spike ascribed to the enhanced horizontal motion. This stage is intermediary to the free-falling (quadratic time dependence) stages  of the spike. Notwithstanding, the experimental investigations \citep{Waddell2001} at low density ratios show the ``non-linear saturation" of spike and bubble to a terminal velocity, for both miscible and immiscible systems. \cite{Goncharov2002} deployed potential theory to extend the ``non-linear saturation" to arbitrary density ratios. In contradiction, \citet{Ramprabhu2006} discovered the development of a reacceleration phase of bubble growth at late time stages, subsequent to the non-linear saturation, for fluids with relatively low to moderate density differences. \citet{Wilkinson2007} substantiated the reacceleration phase experimentally for miscible fluids (with density ratio $\approx 1.35$). The reacceleration phase is attributed to the emanation of the secondary instability in form of Kelvin-Helmholtz (K-H) vortex rings. The K-H rolls reduces the frictional drag between the bubble and spike and engenders a fluid motion that creates a vertical jet of momentum to propel the bubble forward. The chaotic mixing at sufficiently large times disintegrate these K-H rolls \citep{Ramprabhu2012} leading to culmination of reacceleration phase. Continuing the analysis at aforementioned density ratio ($ \approx 1.35$), \citet{Hu2019} numerically investigated the impact of viscosity.  While the study concurred with \citet{Ramprabhu2012} for low-viscosity fluids, it
 revealed a novel "deceleration-acceleration" phenomenon in fluids with moderate viscosity. The ``deceleration-acceleration'' phase is attributed to the transportation of successive K-H rolls to the bubble head prior to the chaotic mixing regime. \citet{Lyubimova2019} investigated the RT instability in a confined domain for miscible fluids exhibiting a temperature sensitive miscibility gap. Such fluids are also referred to as binary fluids. Binary fluids can be categorised on the basis of evolution of miscibility gap as solutions having lower critical solution temperature (LCST) or upper critical solution temperature (UCST). The UCST is the temperature limit beyond which the two fluids are completely miscible, while exhibiting a miscibility gap below it. The capillary effects therefore possess a dynamic nature allowing the use of temperature manipulation to achieve control over flow features. Consequently, several applications such as targeted drug delivery \citep{Schmaljohann2006}, extraction and separation of bio-active compounds (\citet{Kohno2011};\citet{Ventura2017}), 
 enhancement of liquid-liquid mass transfer  \citep{Fornerod2020}, liquid crystal micro-droplet formation \citep{Patel2021} have been reported. 
 \\
 Despite the continued interest, to the best of our knowledge, previous attempts fall short of quantifying the effects of temperature in the context of RT instability in binary fluids. In our previous work \citep{Bestehorn2021}, we deployed a numerical model based on phase-field approach to capture the continuous isothermal transition from immiscible state to miscible state of a binary mixture. The model was further extended \citep{Borcia2022} to account for temperature-field evolution during the phase transition. Herein, we propose an improvement to our previous model to accurately capture the Korteweg stresses beyond the UCST as long as the concentration gradient is non-trivial. The model is employed, to conduct a linear stability analysis for inviscid fluids and extend the dispersion relation \citep{Chandrasekhar1961} for incorporating the effect of partial miscibility between the two fluids. The analysis, utilizing the Boussinesq approximation, demonstrate the early time stage perturbation growth characteristics as a function of the degree of miscibility between the two fluids. Subsequently, a non-linear viscous two-dimensional numerical simulation, relaxing the Boussinesq approximation, is performed to explore the late-time dynamics of RT mixing. 
 \\
 The rest of the paper is organized as follows: Sec. \ref{sec:PFM} describes the novel phase-field model which is employed to reformulate the reformulate the governing equations in Sec. \ref{sec:sysConfig}, wherein, all the dimensional variables are accented by tilde $(\sim)$ sign. The dispersion relation obtained through the linear stability analysis is described in Sec. \ref{sec:LinearAnalysis}. Sec. \ref{sec:numericalFormulation} elucidates the numerical methodology employed to obtain the results discussed in Sec. \ref{sec:resultsDiscussion}. In Sec. \ref{sec:conclusion} concluding remarks are provided. Finally, appendices \ref{appA} and \ref{appB} disseminate supplementary information not contained in the text.

\section{Phase-field modeling: a new approach}
\label{sec:PFM}

The phase-field method \citep{Cahn1958} is employed to capture the interface between the two fluids as it evolves with the flow. The interface is modelled as a thin transition zone with finite thickness governed by the thermodynamic equilibrium state of the fluid pair. The thermophysical properties vary across this transition zone rapidly but smoothly thereby facilitating the definition of an indicator function to numerically pose the problem in one-fluid formulation. An order parameter, $c \in [-1,1]$, is defined to accommodate the spatio-temporal changes of an intensive variable due to the inhomogeneity. The order parameter is employed to define a free energy functional entailing the bulk free energy of respective phases and the excess energy ascribed to the presence of the interface. The free energy functional can be expressed as (\citet{Yue2004}; \citet{Ding2007}):

\begin{equation}
 \tilde{F}(c, \bnabla c) = \int_{\tilde{\Omega}} [\tilde{f_{o}}(c) + \tilde{{\Lambda \over 2}}|{\tilde{\bnabla} c}|^2] d\tilde{\Omega}, 
 \label{freeEnergyFunct}
\end{equation}

\begin{figure}
\centering 
\includegraphics[width=0.6\textwidth]{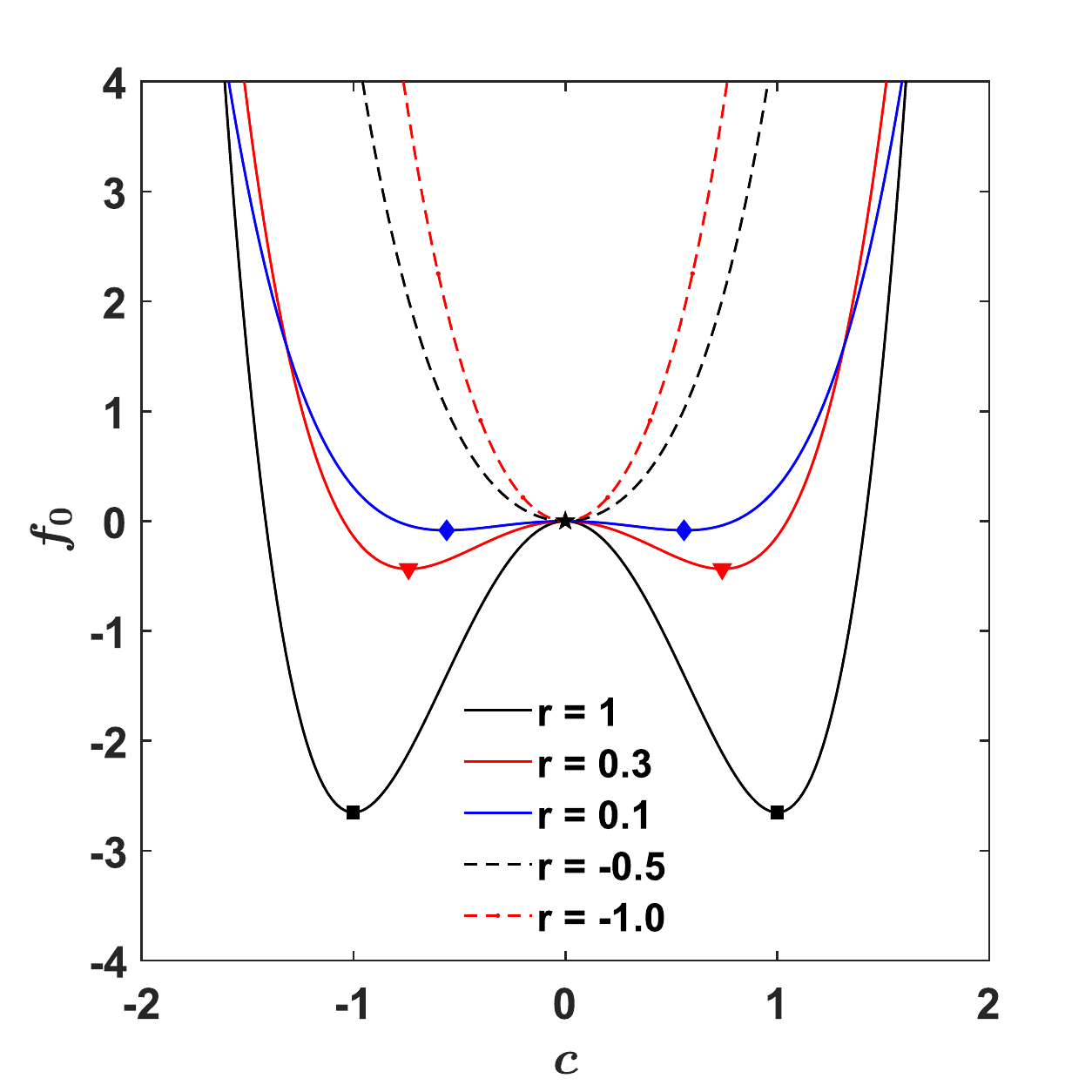} 
\caption{Bulk free energy: Transition from partially miscible state ($r>0$) to miscible state ($r<0$).}
\label{fig1}
\end{figure}

where $\tilde{f_{o}}(c)$ is the bulk free energy density and $\tilde{\Lambda}$ is the mixing energy density. The bulk free energy determines the preferred state of the system in thermodynamic equilibrium as shown in fig. \ref{fig1}. If the bulk free energy has a form of double-well potential \citep{Ginzburg1950}, the system prefers phase-segregation into states corresponding to minimum bulk free energy. Exploiting this feature, \citet{Vorobev2010} introduced empirically defined constants in the expression of the bulk free energy to transform the double-well potential to single-well potential describing the continuous transition from two-phase solution to single-phase mixture in case of binary fluids with UCST. In our previous work, we explicitly linked these constants to the temperature of the system  (\citet{Bestehorn2021};  \citet{Borcia2022}). We employ a bulk free energy formulation motivated from the above-mentioned studies, which can be expressed as:

\begin{equation}
 \tilde{f_{0}}(c) = {\tilde{\Lambda} \over \tilde{\epsilon}^2}({1 \over 4}{|r|^a}{c^4} - {1 \over 2}r{c^2}), 
 \label{bulkFreeEnergy}
\end{equation}
\\
where $\tilde{\epsilon}$ is the capillary width representing the thickness of the interface, with  

\begin{equation}
 r = {\tilde{T_{c}} - \tilde{T} \over \tilde{T_{c}}}, 
 \label{misicbCoeff}
\end{equation}

where $\tilde{T_{c}}$ is the UCST, ``$a$" is an empirical coefficient to be determined experimentally for a given pair of fluids. The dimensionless parameter, $r$, allows to model the continuous transition from partially miscible state ($r>0$) to miscible state ($r<0$). Further, the assumptions made in \citet{Bestehorn2021} and \citet{Borcia2022} were relaxed to accurately account for the Korteweg stresses. The Korteweg stresses \citep{Joseph1990} are induced by large composition gradients and mimic the effects of surface tension. The surface tension $\tilde{\sigma}$, defined as the excess free energy per unit surface area \citep{Cahn1959b}, for a one-dimensional interface is given as \citep{Yue2004}:
 
\begin{equation}
 \tilde{\sigma} = \int_{-\infty}^{\infty} [\tilde{f_{o}}(c) + {\tilde{\Lambda} \over 2}|{\tilde{\bnabla} c}|^2] d\tilde{y}, 
 \label{surfaceTensionTheory}
\end{equation}

The diffuse interface evolves towards the state of minimization of the free energy functional. This state is referred to as the state of thermodynamic equilibrium, characterized by a zero chemical potential. The chemical potential is defined as the variational derivative of the free energy functional given as: 

\begin{equation}
 \tilde{\phi} = {\delta \tilde{F} \over \delta c} = \tilde{f_{0}^\prime}(c) - \tilde{\Lambda} {\tilde{\nabla}^2}c, 
 \label{chemicalPotential}
\end{equation}

Considering an inhomogeneous state $(r > 0)$ at equilibrium ($\tilde{\phi} = 0$), the equation (\ref{chemicalPotential}) can easily be solved (as shown in Appendix \ref{appA}) to obtain the profile of a planar interface, along the gravitational direction $\tilde{y}$, as (\citet{Cahn1958}; \citet{Jacqmin1999}; \citet{Yue2004}; \citet{Bestehorn2021}):

\begin{equation}
 c = - r^{(1-a) \over 2} \tanh\Bigl({{\tilde{y}-\tilde{y_{0}} \over {\sqrt{2} (\tilde{\epsilon} / \sqrt{r}) }}}\Bigr), 
 \label{equilibInterfaceProfile}
\end{equation}

where $\tilde{y_{0}}$ is the location of the interface. The interfacial thickness diverges as $r$ reduces. Further continuing within the purview of thermodynamic equilibrium, the surface tension, $\tilde{\sigma}$, is obtained as (see Appendix \ref{appA}): 

\begin{equation}
 \tilde{\sigma} = {{2 \sqrt{2}} \over 3} {\tilde{\Lambda} \over \tilde{\epsilon}} r^{(3-2a) \over 2} 
 \label{surfaceTensionExpression}
\end{equation}

Eq. \ref{surfaceTensionExpression} can be rewritten as: 

\begin{equation}
 \tilde{\sigma} = {\tilde{\sigma_{0}}}r^{(3-2a) \over 2} = {\tilde{\sigma_{0}}}\bigl({\tilde{T_{c}} - \tilde{T} \over \tilde{T_{c}}}\bigr)^{(3-2a) \over 2} , 
\label{ST_empiricalCorrelation}
\end{equation}

where $\tilde{\sigma_{0}}$ is the surface tension at the limit of immiscibility i.e. $r \to 1$. This expression can then be used to determine the value of the empirical constant $a$. As the temperature of fluid pair approaches the UCST, the extent of mixing permissible for thermodynamic equilibrium rises. Consequently, the concentration gradient decreases thereby reducing the capillarity due to Korteweg stresses. Thus, the constant $a$ must satisfy the two conditions, i.e. $0 < a < 1$ (from Eqn. \ref{equilibInterfaceProfile}) and $a < 3/2$ (from Eqn. \ref{surfaceTensionExpression}). For instance, \citet{May1991} reported the relation, $\tilde{\sigma} = {\tilde{\sigma_{0}}}{[(\tilde{T_{c}}-\tilde{T})/\tilde{T_{c}}]}^{(1.23)}$ for a mixture of isobutyric acid and water (IBW) and therefore, $a_{IBW} = 0.27$.

\begin{figure}
\centering 
\includegraphics[width=0.65\textwidth]{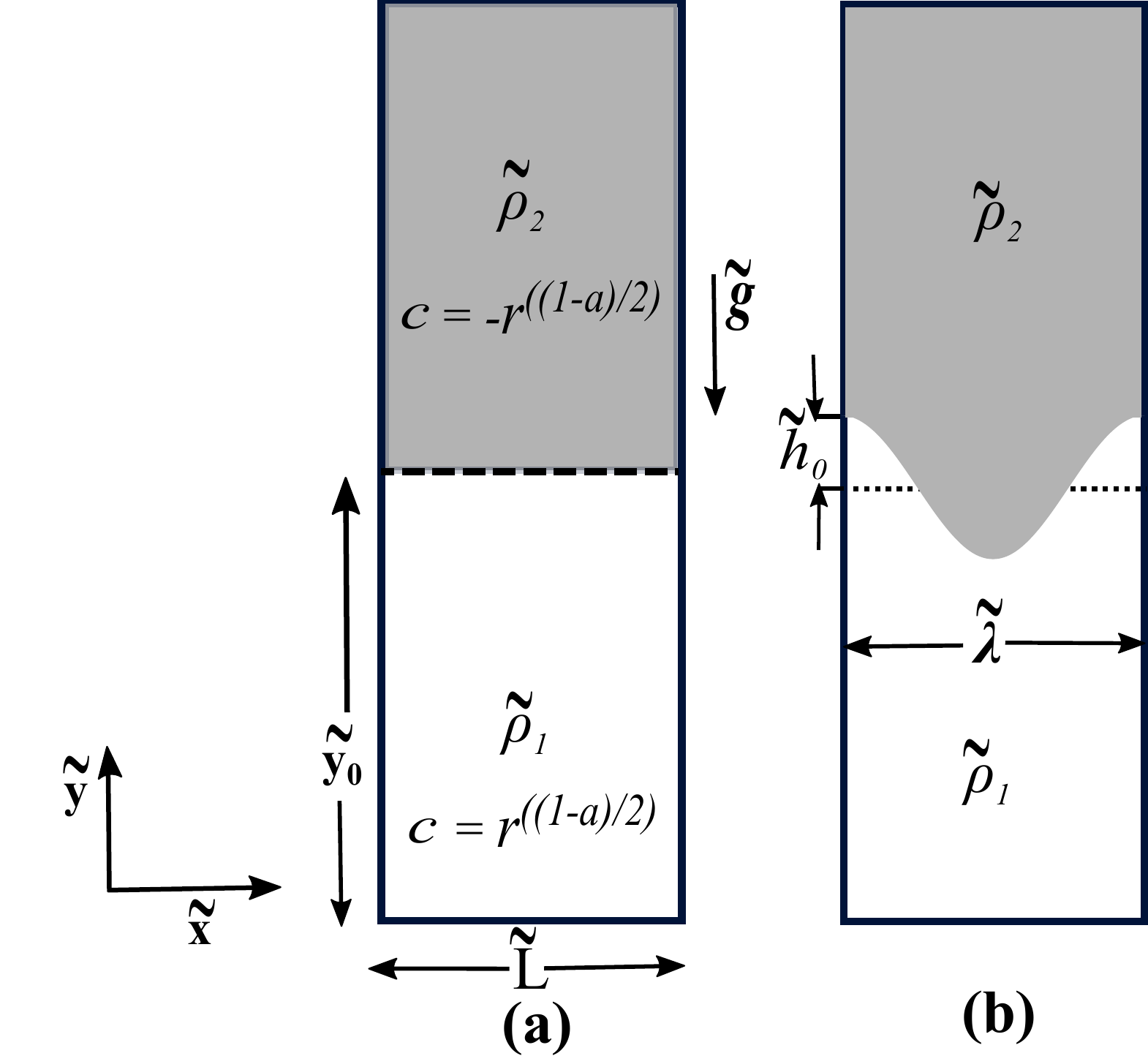} 
\caption{Schematic representation of fluids configuration with heavier fluid of density $\rho_2$ placed over a lighter fluid of density $\rho_1$ in (a) unperturbed state and (b) perturbed state}
\label{fig2}
\end{figure}

\section{System configuration and the governing equations}
\label{sec:sysConfig}

A system comprising of binary fluids with different densities $\tilde{\rho_1}$ and $\tilde{\rho_2}$ is considered with the heavier fluid of density $\tilde{\rho_2}$ superposed over the lighter fluid of density $\tilde{\rho_1}$ in the gravitational field, $\tilde{\textbf{g}}$, as shown in Fig. \ref{fig2}. The equilibrium profile for the planar interface, i.e., the unperturbed interface (see Fig. \ref{fig2}(a)) is obtained by eq. \ref{equilibInterfaceProfile}. A periodic perturbation with amplitude, $\tilde{h_{0}}$, and wavelength, $\tilde{\lambda}$, is imposed on the interface (see Fig. \ref{fig2}(b)) to provoke the RT instability. The wavelength of the perturbation is equal to the lateral dimension $\tilde{L}$ of the domain, with the latter being used as the reference length in the current study. Thus, a profile described by the following dimensionless equation is used as the initial condition:     

\begin{equation}
 c = - r^{(1-a) \over 2} \tanh\Bigl({{y-y_{0} - h_{0}cos(kx) \over {\sqrt{2} {(\Cahn / \sqrt{r})}}}}\Bigr), \label{perturbedInterfaceProfile}
\end{equation}

where $k (= {2\pi \over \lambda})$ is the wavenumber of the imposed perturbation with $\lambda(=\tilde{\lambda}/\tilde{L})$ being the dimensionless wavelength and $\Cahn(= \tilde{\epsilon}/\tilde{L})$ is the Cahn number. The displacement of the phase field order parameter disturbs the initial configuration into a state of non-equilibrium $(\tilde{\phi} \neq 0)$, thereby, emanating a flow towards the state of equilibrium. The evolution of the order parameter, coupled with the hydrodynamic flow, is governed by the advected form of the Cahn-Hilliard (CH) equation \citep{Jacqmin1999}. Employing a reference velocity scale based on acceleration due to gravity ($\tilde{\textbf{u}}_{ref} = \sqrt{\tilde{\textbf{g}}\tilde{L}}$), the dimensionless CH equation as a function of the miscibility parameter $r$ can be written as: 

\begin{equation}
{\partial c \over \partial t}+ \bnabla \bcdot (\textbf{u}c) = {{\Mob} \over \Cahn}{\nabla^2} \Bigl({|r|^a}c^3 - rc - Cn^2 {\nabla^2}c\Bigr), 
\label{CHEquation}
\end{equation}

where $t$ denotes time and $\Mob (= {3 \over {2 \sqrt{2}}} \bigl({\tilde{\sigma_{0}} \over {\tilde{\textbf{u}}_{ref}}\tilde{L}^2}\bigr) \tilde{\gamma} )$ is the dimensionless measure of the mobility $\tilde{\gamma}$ of the interface. An optimum value of mobility is required to avoid, on one hand the shear-thinning of the interface, while on the other hand the over-damping of the flow induced due to high diffusion \citep{Magaletti2013}. A volume-averaged velocity field $\textbf{u}$ \citep{Bagheri2022} is defined to reformulate the Navier-Stokes (NS) equations as a function of the miscibility parameter $r$. The interfacial tension owing to the presence of concentration gradient (Korteweg stresses) is incorporated into the momentum conservation equation following the works of \citet{Jacqmin1999} and \citet{Yue2004}. The equations of motion governing the velocity $(\textbf{u})$ and pressure $(P)$ field can be written as: 

\begin{equation}
\bnabla \bcdot \textbf{u} = 0, 
\label{continuityEquationNumericalModel}
\end{equation}

\begin{eqnarray}
{\partial {\rho_{c}\textbf{u}} \over \partial t}+ \bnabla \bcdot (\rho_{c}\textbf{u} (\times) \textbf{u}) & = & -\bnabla {P} + {1 \over \Rey}[\bnabla \bcdot \bigl(\mu_{c}(\bnabla \textbf{u} + {(\bnabla \textbf{u})}^{T})\bigr)] + \rho_{c}\textbf{g}
 \nonumber\\
 && \mbox{} + {{3 \over {2 \sqrt{2}}}{1 \over {\Web\Cahn}}} \Bigl({|r|^a}c^3 - rc - \Cahn^2 {\nabla^2}c\Bigr)\bnabla c, 
\label{momentumEquationNumericalModel}
\end{eqnarray}

where $\Rey(= {{\tilde{\rho}_{1} \tilde{\textbf{u}}_{ref} \tilde{L}} \over \tilde{\mu}_{1}})$ is the Reynolds number and $\Web(= { {\tilde{\rho}_{1} {\tilde{\textbf{u}}_{ref}}^2 \tilde{L}}  \over \tilde{\sigma}_{0}})$ is the Weber number.
Further, $\rho_{c}$ and $\mu_{c}$ are order parameter based dimensionless density and dynamic viscosity, which can be written as: 

\begin{equation}
\rho_{c} = {\tilde{\rho} \over \tilde{\rho}_{1}}= \Bigl({1 + c \over 2}\Bigr) + {\tilde{\rho}_{2} \over \tilde{\rho}_{1}} \Bigl({1 - c \over 2}\Bigr), 
\label{volumetricAverageDensity}
\end{equation}

\begin{equation}
\mu_{c} = {\tilde{\mu} \over \tilde{\mu}_{1}}= \Bigl({1 + c \over 2}\Bigr) + {\tilde{\mu}_{2} \over \tilde{\mu}_{1}} \Bigl({1 - c \over 2}\Bigr), 
\label{volumetricAverageDynamicViscosity}
\end{equation}

with $\mu_{1}$ and $\mu_{2}$ being the dynamic viscosity of the lighter and the heavier fluid respectively.

\section{Linear stability analysis}
\label{sec:LinearAnalysis}
We employed the methodology proposed by \cite{Celani2009} to reformulate the well-known dispersion relation for RT instability \citep{Chandrasekhar1961} in the context of the binary fluids as the degree of partial miscibility is varied. The CH equation (eq. \ref{CHEquation}) is coupled with the inviscid form of the NS equations under Boussinesq approximation \citep{Kundu2001}.  The equation of motion governing the velocity $(\textbf{u})$ and pressure $(P)$ field for a potential flow can be written as:

\begin{eqnarray}
{\partial \textbf{u} \over \partial t}+ \textbf{u} \bcdot \bnabla  \textbf{u} & = & -\bnabla P - \textit{A} c \textbf{g}
 \nonumber\\
 && \mbox{}  + {{3 \over {2 \sqrt{2}}}{1 \over {\Web_{B}\Cahn}}} \Bigl({|r|^a}c^3 - rc - \Cahn^2 {\nabla^2}c\Bigr)\bnabla c, 
\label{BoussinesqMomentumEquation}
\end{eqnarray}

where $\textit{A} (={\tilde{\rho}_{2}-\tilde{\rho}_{1} \over \tilde{\rho}_{1} + \tilde{\rho}_{2}})$ is the Atwood number, $\Web_{B} (= {\tilde{\rho_{0}} {{\tilde{\textbf{u}}_{ref}}^2} \tilde{L} \over \tilde{\sigma_{0}}})$ is the Boussinesq Weber number defined employing the mean density $\tilde{\rho_{0}}( = {\tilde{\rho}_{1}+ \tilde{\rho}_{2} \over 2} )$. 

For small amplitude perturbations, the eq. \ref{BoussinesqMomentumEquation} can be linearized and subsequently integrated following the work of \citet{Celani2009} to obtain the dispersion relation for gravity-capillary waves as function of the miscibility parameter $r$ as follows (see Appendix \ref{appB} and supplementary material):

\begin{equation}
\omega^2(k) = {k^3 \over 2\Web_{B}}r^{(3-2a) \over 2} + k\textit{A}gr^{(1-a) \over 2}, 
\label{dispersionRelationGCW}
\end{equation}

where $\omega$ is the frequency of oscillation of the perturbed interface about the mean position. The equation reduces to the classical dispersion relation \citep{Chandrasekhar1961} in the limit of immiscibility $(r \to 1)$. Considering a pair of fluids with unstable density stratification, the growth rate $\alpha(k)$ of the perturbation for RT instability is therefore given as: 

\begin{equation}
\alpha^2(k) = k\textit{A}gr^{(1-a) \over 2} - {k^3 \over 2\Web_{B}}r^{(3-2a) \over 2}, 
\label{dispersionRelationRTI}
\end{equation}

Furthermore, for a given combination of wavenumber $k$ and Boussinesq Weber number $We_{B}$, the threshold value of the parameter $r$, for the stability of the partially diffuse interface to small perturbations, can be obtained in straightforward manner $(\alpha(k) = 0)$ to be:  

\begin{equation}
r_{th} = \Bigl({2\Web_{B}\textit{A}g  \over k^2  }\Bigr)^{({2 \over 2-a})}, 
\label{cutoffMiscibCoeff}
\end{equation}

The Boussinesq approximation employed herein renders the above-mentioned expressions applicable to fluid pairs with small density differences. In order to complete the analysis, we performed numerical investigations by relaxing the Boussinesq approximation as described in Sec. \ref{sec:numericalFormulation}.

\section{Numerical Formulation}
\label{sec:numericalFormulation}

We investigate the dynamics of the single-mode RT instability in following two configurations:\\ $(\mathrm{1})$ the fluid pair is assumed to be at the thermodynamic equilibrium, as determined from the temperature of the system, prior to perturbing the interface,  \\ $(\mathrm{2})$ the fluids are just brought into contact and therefore are in a state of thermodynamic non-equilibrium when the interface is perturbed. \\ The isothermal early-stage growth characteristics and the late-time interface evolution is studied, as a function of the  proximity of distinct system temperatures to the UCST. Fig. \ref{fig2}(b) depicts the schematic representation of the computational domain. The size of the rectangular domain is $[L \times 6L]$, unless stated otherwise. The numerical simulations are performed in Cartesian framework, with finite volume formulation on a uniform mesh. The fluids are assumed to be Newtonian and incompressible. The thermophysical properties of the bulk phases evolve as a function of phase field order parameter as a consequence of mixing due to interfacial diffusion.

\subsection {Initial and boundary conditions}

At the start of the computation (at $t = 0$), the two fluids with equal volume are assumed to be stationary. For simulations pertaining to configuration $(\mathrm{1})$, distinct values of $r$ are considered while initializing the order parameter profile by employing eq. \ref{perturbedInterfaceProfile}. This value of $r$ is then held constant throughout the process of interfacial evolution. Whereas, the configuration $(\mathrm{2})$ is analyzed by initializing the interface profile using $r = 1$ in eq. \ref{perturbedInterfaceProfile} as the fluids are just brought into contact. Subsequently, two distinct values of $r$ are considered to study the combined effect of interfacial diffusion and hydrodynamic flow as the system is driven towards the state of equilibrium. The geometrical parameters pertinent to initialization are elucidated in ensuing sections. The governing equations (eqns. \ref{CHEquation} - \ref{momentumEquationNumericalModel}) are subjected to periodic (cyclic) boundary conditions in the lateral direction and zero gradient boundary conditions in the vertical direction \citep{Ramprabhu2012}.

\subsection {Numerical methodology}

In the present study, the phaseFieldFoam solver (\citet{Jamshidi2019}; \citet{Bagheri2022})  is extended to solve the coupled CH-NS equations (eqns. \ref{CHEquation} - \ref{momentumEquationNumericalModel}) in a segregated manner. The velocity and the pressure field is coupled by employing the PIMPLE algorithm. The PIMPLE algorithm is a combination of the PISO (Pressure implicit with splitting of operator) \citep{Issa1985} and the SIMPLE (Semi-implicit method for pressure linked equations) \citep{Patankar1972} algorithms. The coupling of the order parameter field with the velocity and pressure fields is achieved by first solving the CH equation (eq. \ref{CHEquation}) at the beginning of each PIMPLE iteration, followed by updating the Korteweg stress term in NS equations (last term of eq. \ref{momentumEquationNumericalModel}) and finally solving the mass and momentum transport equations (eqns. \ref{continuityEquationNumericalModel} and \ref{momentumEquationNumericalModel}). \\
The phaseFieldFoam is based on OpenFOAM which discretizes the governing equations following a second-order finite-volume scheme. Thus, the CH equation is split into two second-order equations during the solution procedure, to facilitate the discretization of the fourth order diffusion term. First, the chemical potential $\phi$ is calculated which is subsequently substituted in the diffusion term of the CH equation to update the order parameter $(c)$ field. \\
The advective terms in the CH equation (eq. \ref{CHEquation}) and the momentum equation (eq. \ref{momentumEquationNumericalModel}) are
discretized using the Gauss Gamma \citep{Jasak1999} and Gauss limited linear V \citep{Sweby1984} schemes, respectively, both of which are high-resolution, nonlinear NVD (Normalized
Variable Diagram) and TVD (Total Variation Diminishing) schemes \citep{Moukalled2016}.
These schemes have been chosen as they are designed to preserve the sharpness of the interface and
prevent numerical oscillations. Meanwhile, the diffusive terms in these equations are
discretized using the Gauss linear method, and the temporal terms are handled using the
first-order implicit Euler time-stepping scheme \citep{Ferziger2002}. Subsequently, preconditioned bi-conjugate gradient (BiCG) method of \citet{VanDerVorst1992} is used to iteratively solve the coupled equations.

\begin{figure}
\centering 
\includegraphics[width=1.0\textwidth]{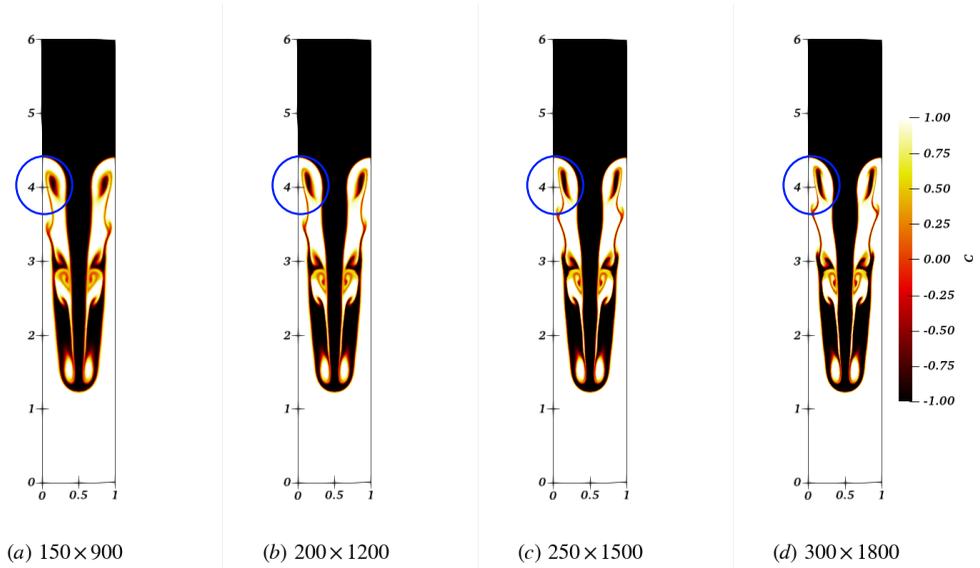} 
\caption{The interface topology at $t = 8.64$ for different grid sizes given by $(a) 150 \times 900$, $(b) 200 \times 1200$, $(c) 250 \times 1500$ and $(d) 300 \times 1800$.}
\label{fig3}
\end{figure}

\begin{figure}
\centering 
\includegraphics[width=1.0\textwidth]{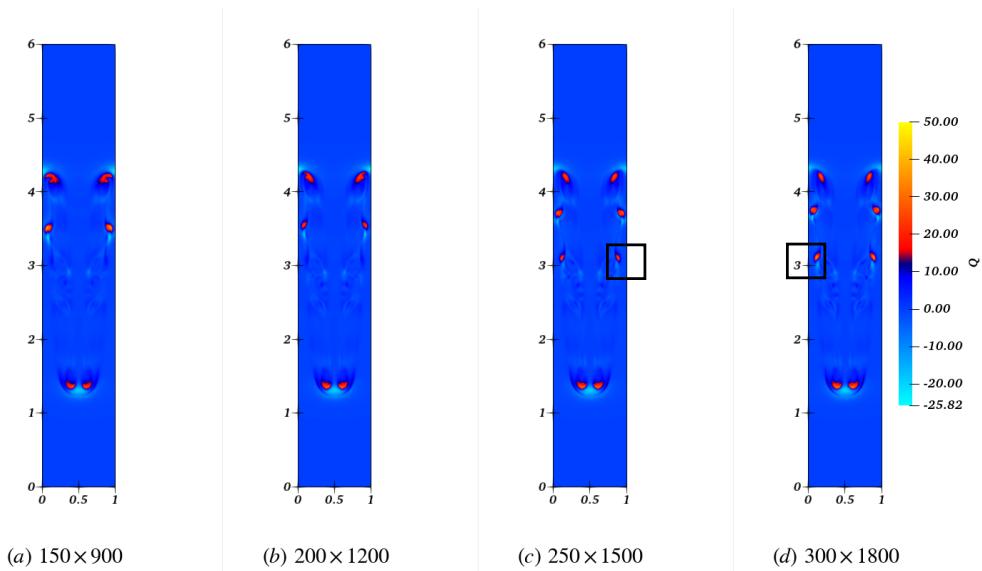} 
\caption{The Q-criteria at $t = 8.64$ for different grid sizes given by $(a) 150 \times 900$, $(b) 200 \times 1200$, $(c) 250 \times 1500$ and $(d) 300 \times 1800$.}
\label{fig4}
\end{figure} 

\subsection {Validation}

\begin{figure}
\centering 
\includegraphics[width=0.4\textwidth]{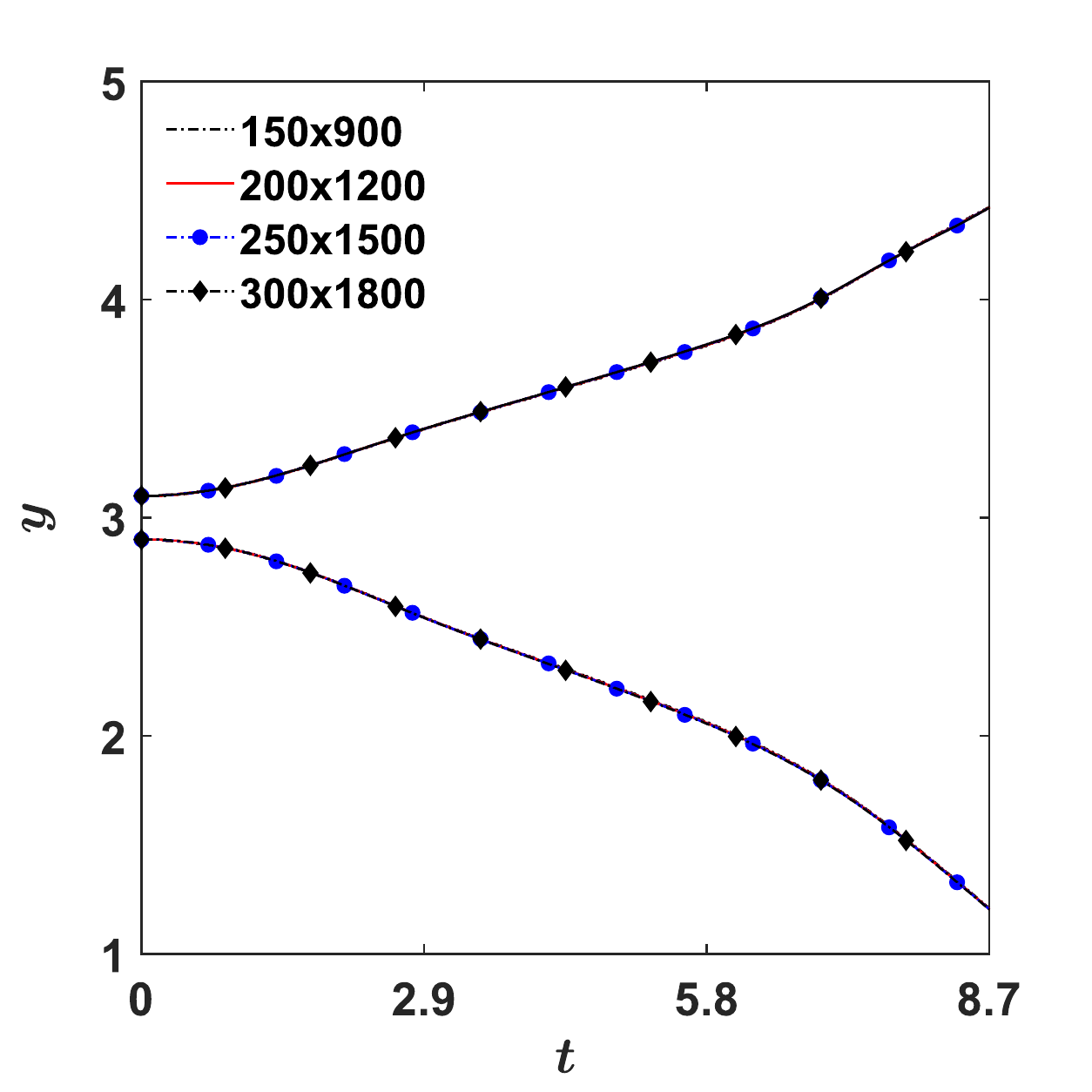} \hspace{0.2cm}
\includegraphics[width=0.4\textwidth]{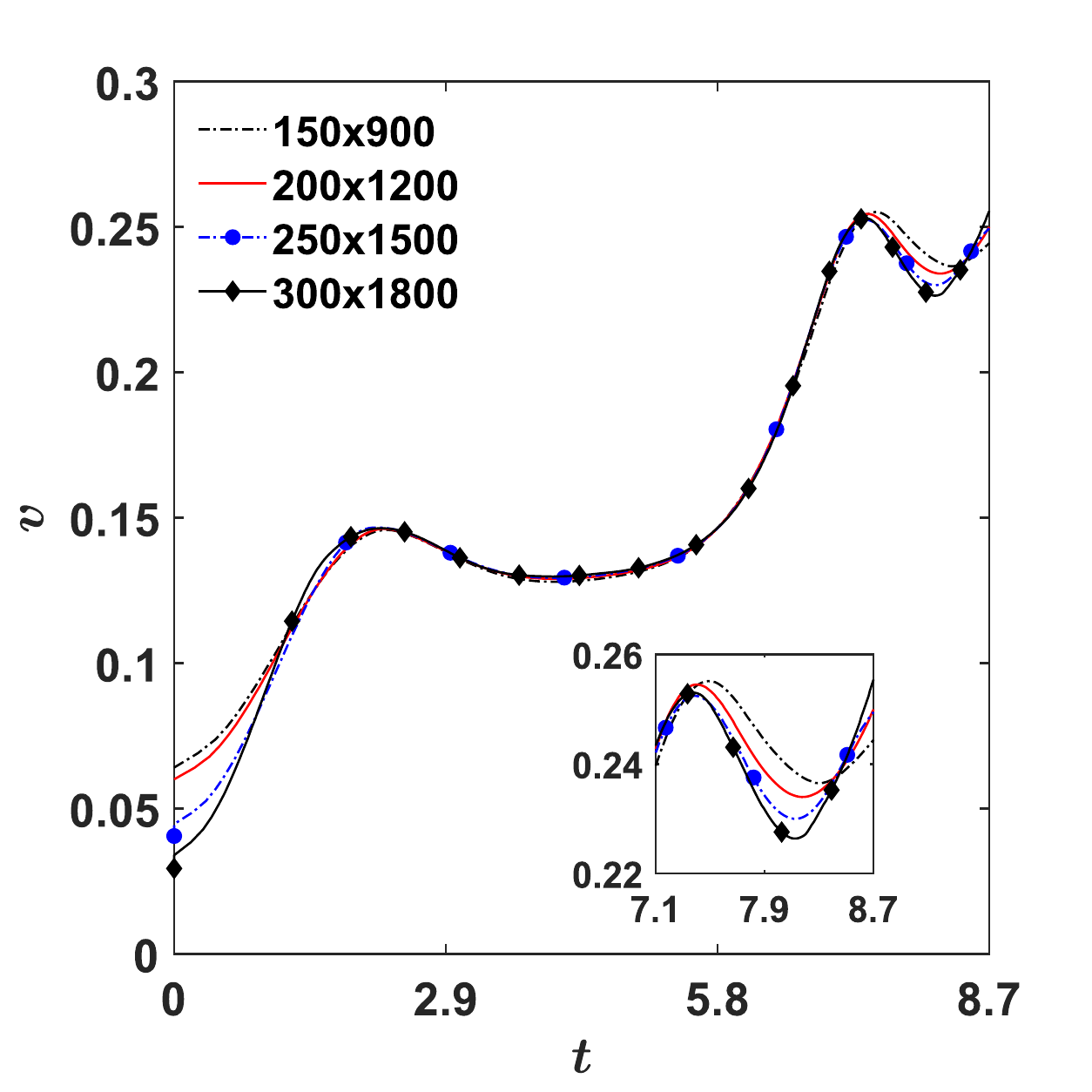} \\
\hspace{0.1cm} (a) \hspace{5.3cm} (b) \\
\caption{Temporal evolution of the (a) spike and bubble tip location and (b) bubble tip velocity for different grid sizes namely $ 150 \times 900$, $200 \times 1200$, $250 \times 1500$ and $300 \times 1800$. The dimensionless parameters used in these calculations are $\textit{A} = 0.2$, $\Rey = 2000$, $\Web = 10^{5}$, $\Cahn = 0.01$.}
\label{fig5}
\end{figure}

 We present the qualitative as well as quantitative comparison of the current numerical solver with the benchmark case described in \citet{Hamzehloo2021}. The order parameter profile governing the fluids configuration is initialized (eq.  \ref{perturbedInterfaceProfile}) with an initial perturbation of amplitude $h_{0} = 0.1$ and $r = 1$. The wavenumber of the perturbation $k(= 2\pi)$ is constant throughout the current study. The thermophysical properties of the fluids are governed through Atwood number, Reynolds number and the Weber number, chosen here to be $\textit{A} = 0.2$, $\Rey = 2000$ and $\Web = 10^{5}$. Furthermore, the $\Cahn(=0.01)$ is held constant throughout the analysis. Consequently, the dimensional mobility $\tilde{\gamma}$ is also maintained constant throughout the analysis with $\tilde{\gamma} = 0.01 \tilde{\epsilon}^2$ \citep{Jamshidi2019}. The dimensionless gravitational acceleration acts downwards with magnitude $g = 1.0$.  To begin with, a grid independence test is conducted to obtain the optimum grid size for further investigations. Four different grid sizes are considered, namely, $150 \times 900$, $200 \times 1200$, $250 \times 1500$ and $300 \times 1800$. The results are presented through fig. \ref{fig3} -  \ref{fig5}. Fig. \ref{fig3} and fig. \ref{fig4} depicts the interface topology and the Q-criteria at $t = 8.64$ respectively. The Q-criteria \citep{Hu2019} is a vortex identification method signifying the excess of local rotation rate over strain rate. The expression of Q-criteria can be obtained by employing Frobenius norm as follows \citep{Epps2017}:

 \begin{equation}
 Q = {1 \over 2}\biggl( {||\Omega||}^{2} - {||S||}^{2} \biggr), 
 \label{Q_Criteria_definition}
 \end{equation}

 where $\Omega$ is the angular rotation rate tensor and $S$ is the strain rate tensor. In a two-dimensional Cartesian framework, eq. \ref{Q_Criteria_definition} can be simplified to give:

 \begin{equation}
 Q = -{1 \over 2}\biggl[ \biggl({\partial u \over \partial x}\biggr)^2 + \biggl({\partial v \over \partial y}\biggr)^2 \biggr] - \biggl({\partial u \over \partial y} \biggr)\biggl({\partial v \over \partial x} \biggr), 
\label{Q_Criteria}
\end{equation}

\begin{figure}
\centering 
\includegraphics[width=0.5\textwidth]{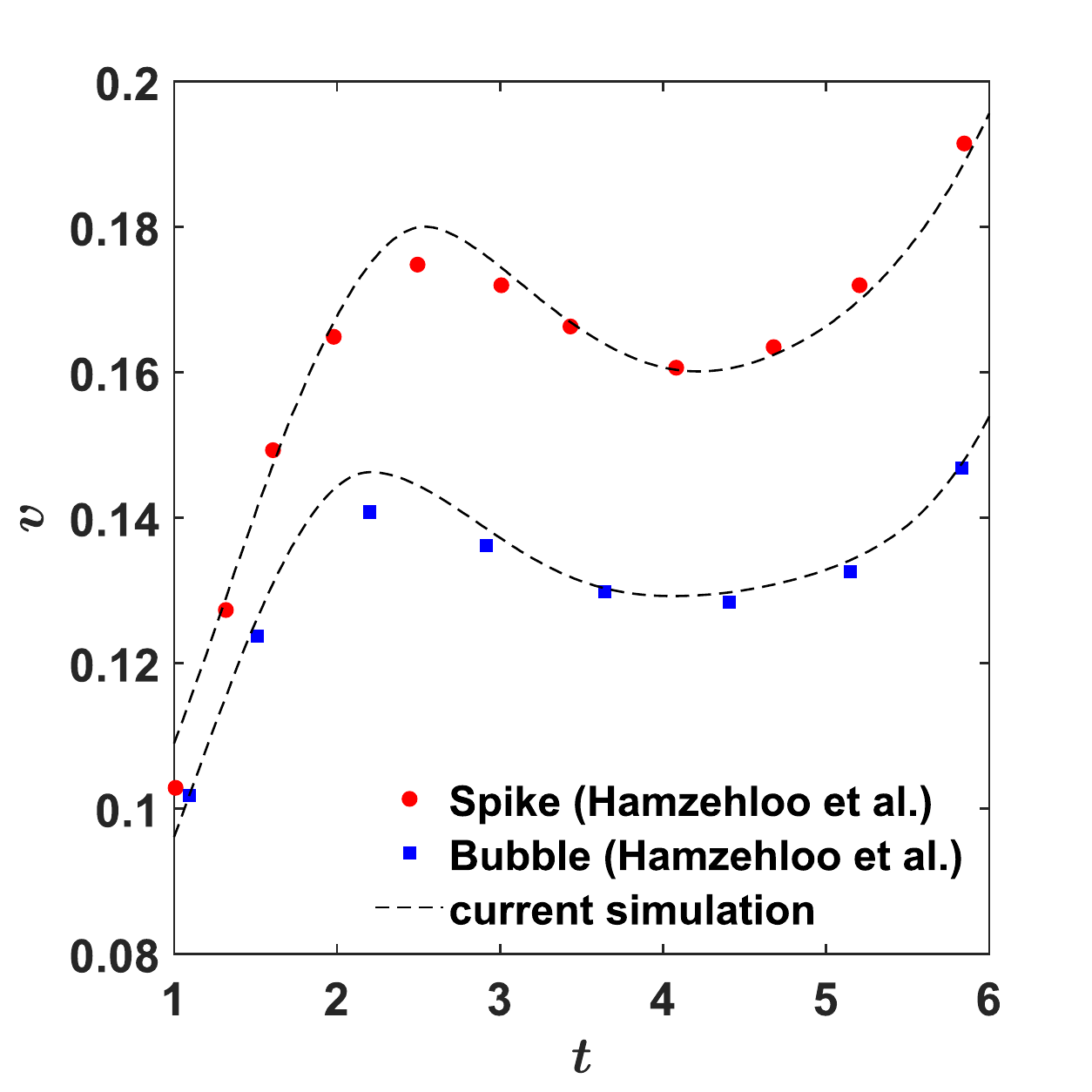} 
\caption{Comparison of the bubble velocity obtained from the current solver with benchmark results \citep{Hamzehloo2021}.}
\label{fig6}
\end{figure}

 The instantaneous location of the tip of the bubble and spike is captured by tracking the iso-lines of $c = 0$ along the axes $x = 0$ and $x = 0.5$ respectively. The early stage temporal evolution of the interface topology was found to be identical in all the cases which is corroborated quantitatively in fig. \ref{fig5}(a), demonstrating the temporal evolution of the location of the tip of rising bubble and the falling spike. However, at late time stages, the formation of vortex-dominated flow regions is well captured in finer grids, namely, $250 \times 1500$ and $300 \times 1800$ as shown in fig. \ref{fig3} and fig. \ref{fig4}. The dissipation of vortices (fig. \ref{fig4}) in $150 \times 900$ and $200 \times 1200$ grids is ascribed to the artificial viscosity arising due to large grid size. The rectangular insets in the case of $250 \times 1500$ and $300 \times 1800$ highlights the ability of finer grids to resolve the secondary flow structures. Consequently, the velocity of the rising bubble is sensitive to the grid size \citep{Ramprabhu2012} as shown in fig. \ref{fig5}(b). Thus,  $250 \times 1500$ is chosen as the optimum grid size for generating the remaining results. Subsequently, we present the comparison of the bubble velocity obtained from the current solver with that of reported by \citet{Hamzehloo2021} in fig. \ref{fig6}. The agreement between the two solutions establishes the accuracy of our numerical solver for forthcoming results.

\section{Results and Discussion}
\label{sec:resultsDiscussion}

Herein, we elucidate the results from the linear stability analysis and the numerical simulations. This section is divided into two subsections, namely, $(\mathrm{1})$ initialization with thermodynamic equilibrium where the fluids are allowed to diffuse according to the system temperature prior to perturbing the interface and $(\mathrm{2})$ initialization with a perturbed interface with non-equilibrium profile. The analysis performed under the purview of configuration $(\mathrm{1})$ aids to reveal the characteristic behavior of binary fluids exhibiting miscibility gap as a function of temperature, as the system approaches the UCST. On the other hand, the second configuration sheds light on the relative dominance of advection due to fluid flow over chemical potential driven diffusion.      

\subsection {Initialization with thermodynamic equilibrium}

We begin our analysis employing the derived dispersion relation for RT instability in the context of binary fluids. The dispersion relation, originally derived under the Boussinesq approximation for inviscid fluids, is given by eq. \ref{dispersionRelationRTI}. The Boussinesq Weber number $(\Web_{B})$ therein, can be replaced by Weber number $(\Web)$ defined in sec. \ref{sec:sysConfig} to rewrite the dispersion relation as follows:   

\begin{equation}
\alpha^2(k) = k\textit{A}gr^{(1-a) \over 2} - {{k^3 (1 - \textit{A})} \over 2\Web} r^{(3-2a) \over 2}, 
\label{dispersionRelationRTI_numericalWe}
\end{equation}

\begin{figure}
\centering 
\includegraphics[width=0.4\textwidth]{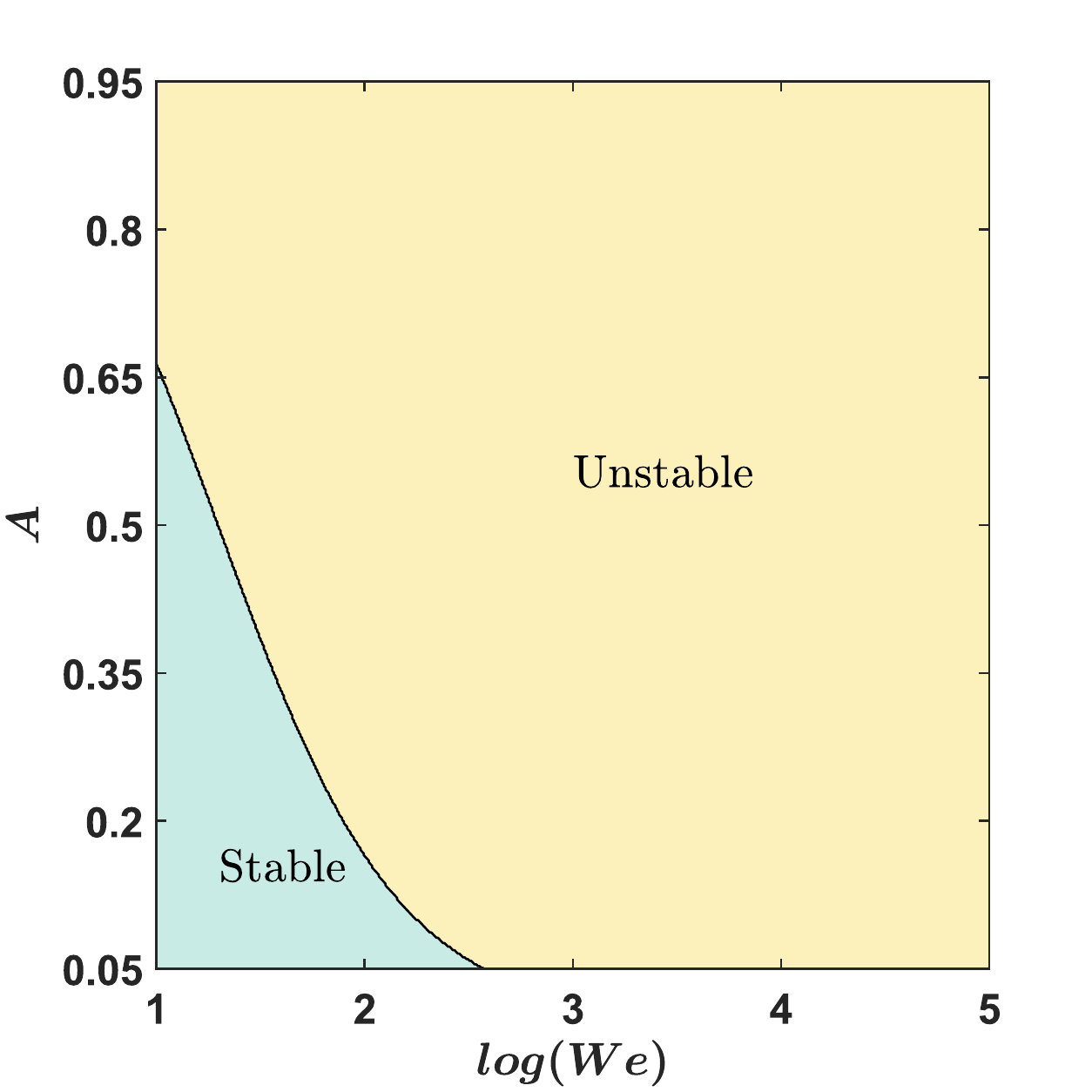} \hspace{0.2cm}
\includegraphics[width=0.4\textwidth]{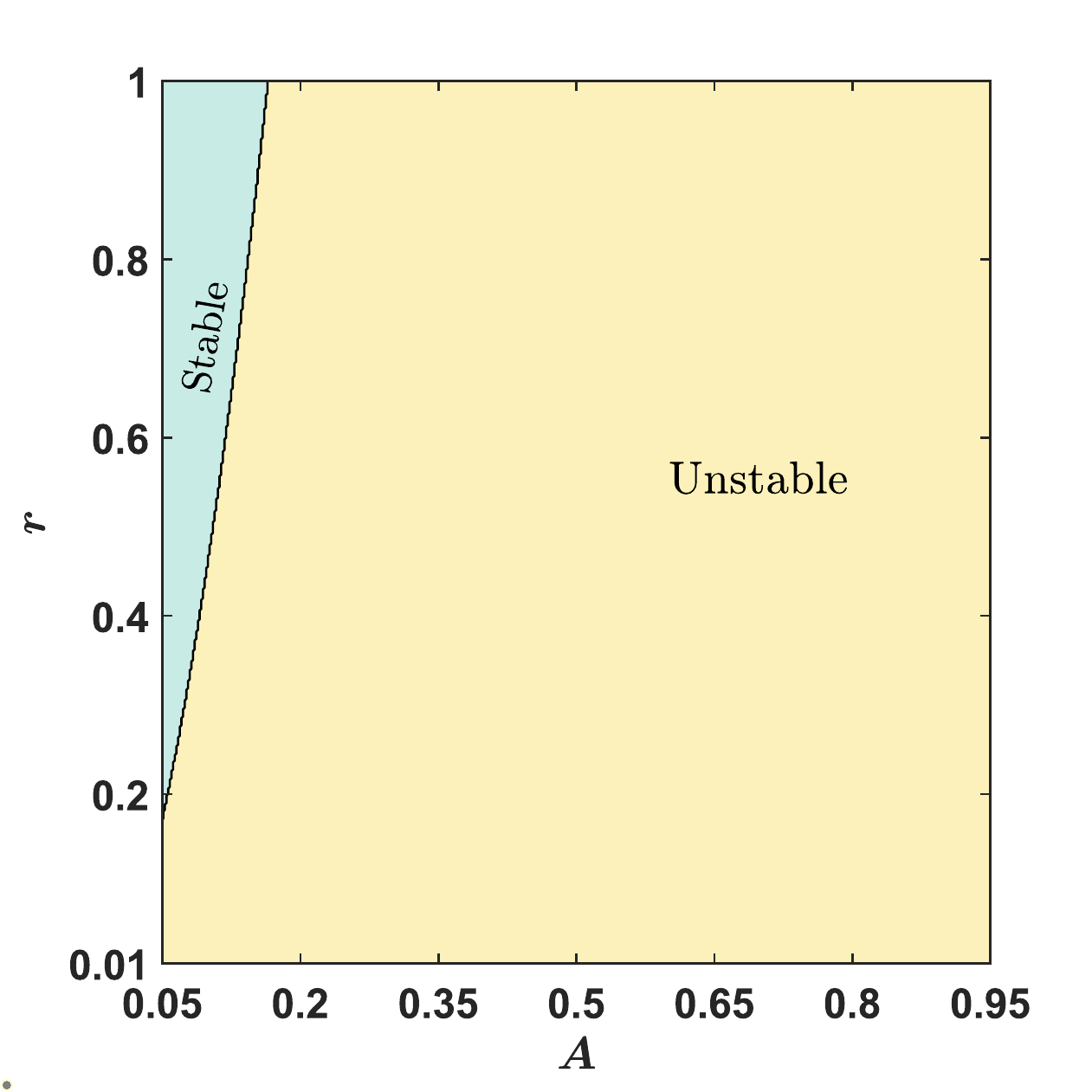} \\
\hspace{0.1cm} (a) \hspace{5.3cm} (b) \\
\includegraphics[width=0.4\textwidth]{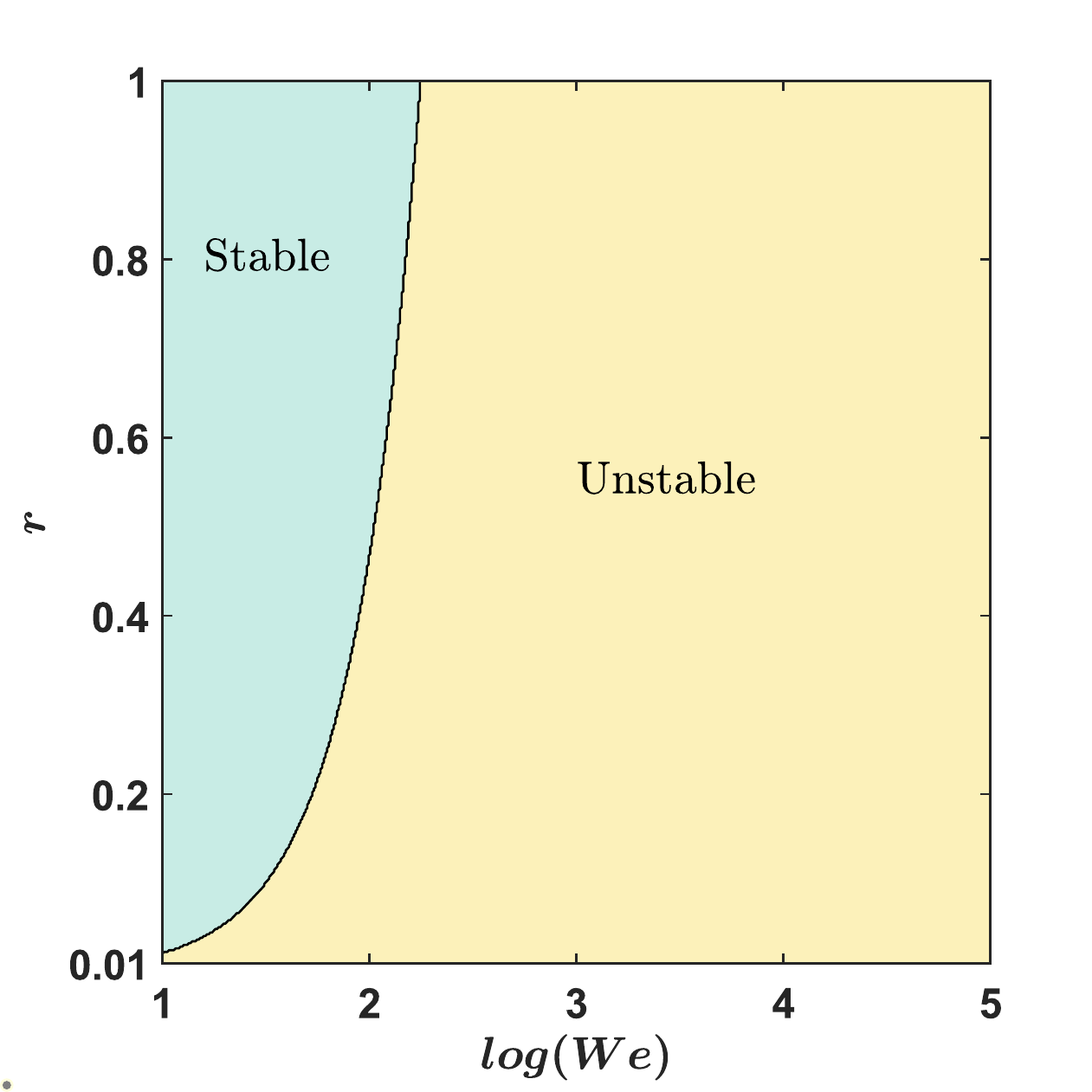} \hspace{0.2cm}
\includegraphics[width=0.4\textwidth]{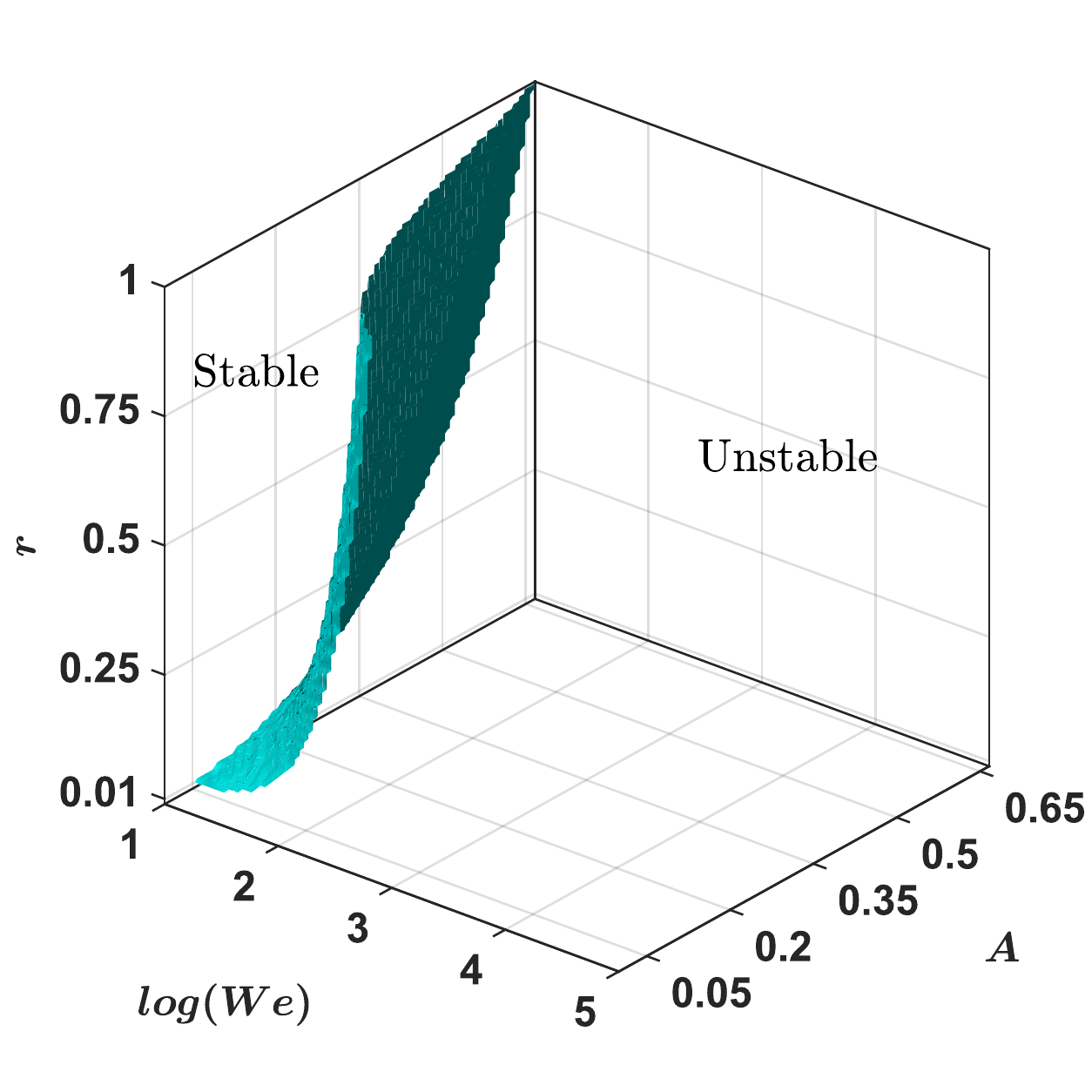} \\
\hspace{0.1cm} (c) \hspace{5.3cm} (d) \\
\caption{Demarcating the boundary between stable and unstable configurations for RT instability in binary fluids through marginal stability curves for (a) $A$ vs $log(\Web)$ at $r = 1$, (b) $r$ vs $A$ at $\Web = 100$, (c) $r$ vs $log(\Web)$ at $A = 0.1$, (d) Three-dimensional representation of the stability boundary }
\label{fig7}
\end{figure}

Eq. \ref{dispersionRelationRTI_numericalWe} is employed to demarcate the boundary between stable and unstable configuration for binary fluids exhibiting temperature sensitive miscibility gap as shown in fig. \ref{fig7}. A parametric study is conducted by varying the density contrast, the surface tension and the proximity to the UCST governed by the $\textit{A}$, $\Web$ and $r$ respectively. The empirical constant $a$ is assumed to be constant (we choose $a = 0.5)$ throughout this study. The continuous approach to the UCST is marked with a monotonous decrease in the value of $r$ (eq. \ref{misicbCoeff}). The associated increase in relative miscibility as $r$ decreases renders the thermophysical properties of the bulk phases as well as the interfacial tension to be dynamically linked. To begin with, we reproduce the behaviour of immiscible fluids $(r = 1)$ as shown in fig. \ref{fig7}(a). The fluid pair is stable to small perturbations if $\textit{A}$ and $\Web$ are low. A relatively lower $\Web$ corresponds to higher surface tension in the limit of immiscibility $(\tilde{\sigma_{0}})$ (re-emphasizing that the $\Web$ is independent of the miscibility parameter $r$ in the current analysis). The high surface tension drives the interface towards the state of minimum possible energy by reducing the surface area i.e. reducing the amplitude of the perturbation imposed thereby stabilizing the fluid pair despite an unstable density stratification. As the $\Web$ is increases, the ability of the interfacial tension to counteract the destabilizing effect of buoyancy ascribed to the density contrast diminishes, thereby continuously expanding the zone of instability. Subsequently the effect of decrease in $r$ on the marginal stability is investigated. Fig. \ref{fig7}(b,c,d) depicts the marginal stability curve for (b) a fluid pair with $\Web = 100$ as a function of $r$ and $\textit{A}$, (c) a fluid pair with $\textit{A} = 0.1$ as a function of $r$ and $\Web$ and (d) three dimensional representation encapsulating the overall mechanism. The regime of instability was found to expand as the fluid system approaches the UCST. 

\begin{figure}
\centering 
\includegraphics[width=0.4\textwidth]{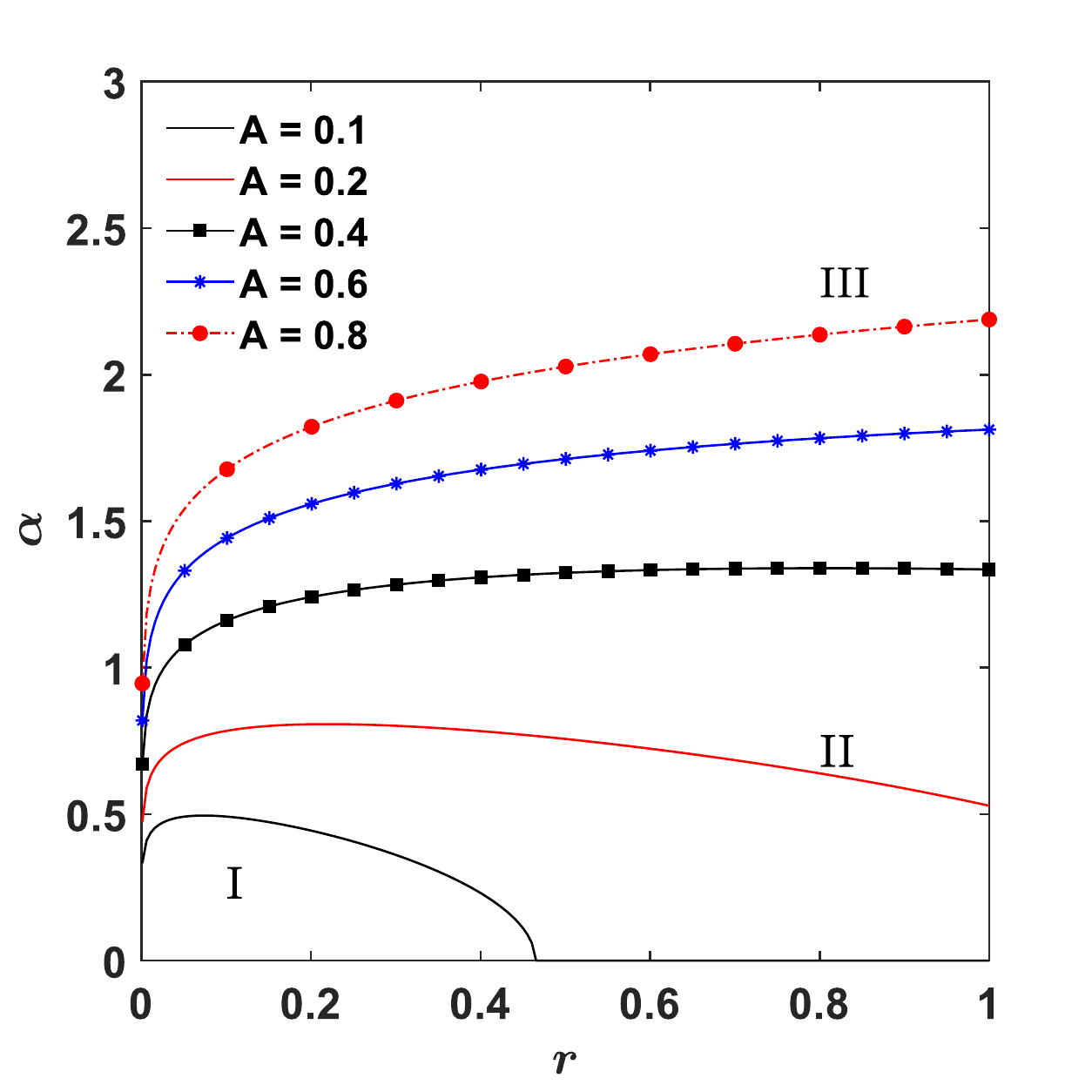} \hspace{0.2cm}
\includegraphics[width=0.4\textwidth]{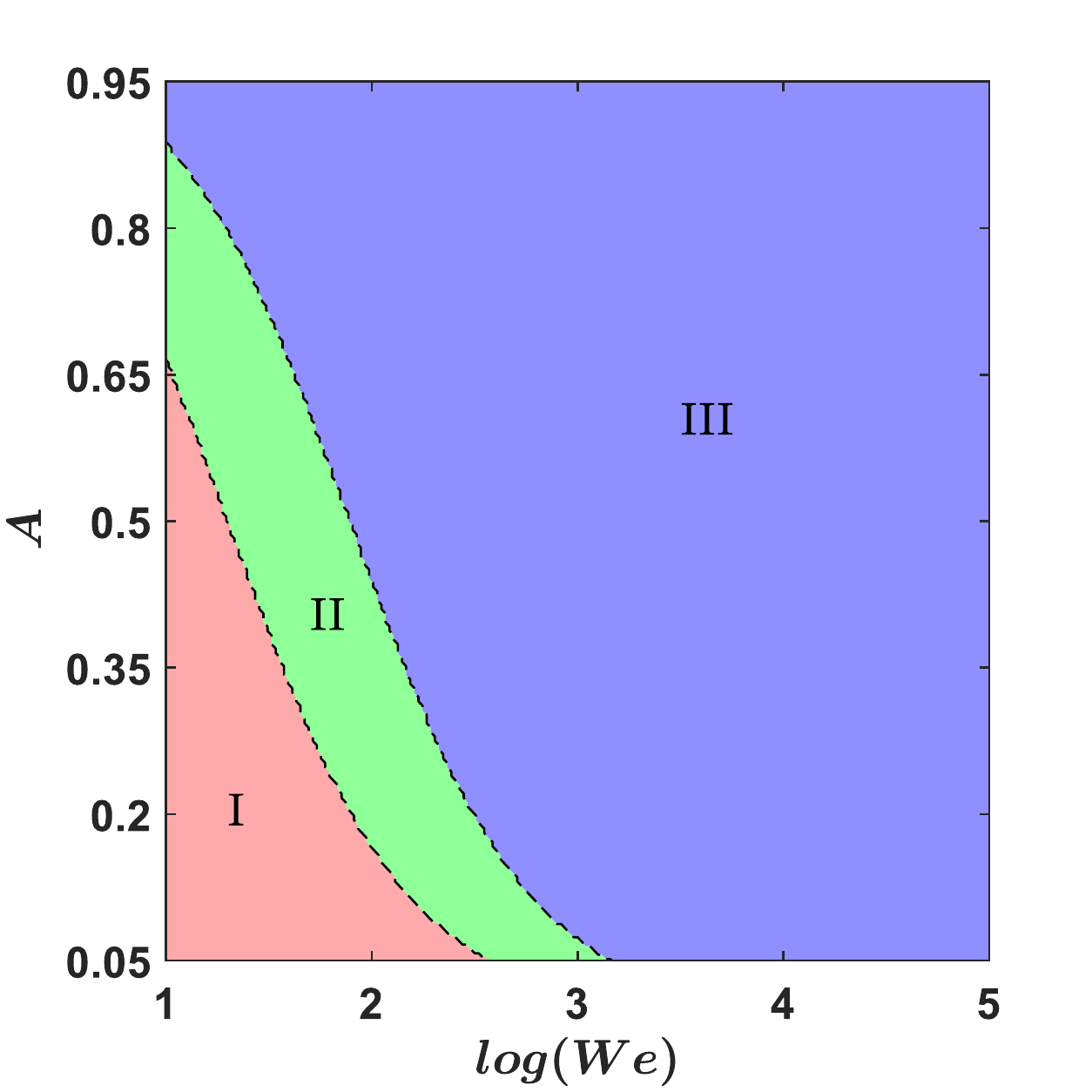} \\
\hspace{0.1cm} (a) \hspace{5.3cm} (b) \\
\caption{(a)Variation of the growth rate $\alpha$ with $r$ at $\Web = 100$ for different values of $A$ enumerated as $A = 0.1, 0.2, 0.4, 0.6$  and $0.8$. (b) A regime map to demarcate the boundaries between different zones of variation of the growth rate $\alpha$ with $r$ }
\label{fig8}
\end{figure}

The decrease in $r$ is associated not only with a decrease in surface tension (eq. \ref{ST_empiricalCorrelation}) but also a decrease in effective density contrast due to the mixing of two fluids. The relative dominance of these two effects, which governs the perturbation growth mechanism at the interface, is quantified utilizing eq. \ref{dispersionRelationRTI_numericalWe}. The variation in growth rate $\alpha$ with $r$ is qualitatively characterized in three distinct zones namely zone $I, II$ and $III$ as shown in fig. \ref{fig8}(a). For a fluid pair with relatively low density contrast and sufficiently high surface tension, at the limit of immiscibility i.e. the system temperature being far from the UCST $(r = 1)$, the interface is stable to small perturbations. As the value of $r$ is decreased, the decrease in surface tension dominates over the decrease in effective density contrast thereby provoking the instability. The threshold value of the miscibility parameter $r$ for transition from the stable state to the unstable state is given by eq. \ref{cutoffMiscibCoeff}, which can be re-written as:                     

\begin{equation}
r_{th} = \Bigl({2\Web\textit{A}g  \over k^2 (1 - \textit{A})  }\Bigr)^{({2 \over 2-a})}, 
\label{cutoffMiscibCoeff_numericalWe}
\end{equation}

The fluid pairs exhibiting this transition behaviour are characterized as zone $I$ binary fluids. Naturally, the threshold miscibility parameter is only defined for binary fluids in zone $I$. As the coefficient $r$ is decreased below the threshold $r_{th}$, the growth rate $\alpha$ increases before saturating to a peak growth rate and subsequently decreases as further decrease in $r$ leads to domination of decrease in density contrast over the decrease in surface tension. Fig. \ref{fig8}(a) demonstrates this behaviour for $\textit{A} = 0.1$ at $\Web = 100$. Further, fluid pairs having moderate to high density contrast and relatively moderate to low surface tension at $r = 1$ are unstable to small perturbations at the interface irrespective of the proximity of the system temperature to UCST. Such fluids are characterized to be either zone $II$ or zone $III$ binary fluids. The growth rate $\alpha$ for zone $II$ binary fluids, as shown by the fluid pair $\textit{A} = 0.2$ and $\Web = 100$ in fig. \ref{fig8}(a), follows similar behaviour as that of fluids in zone $I$. Conversely, for zone $III$ binary fluids, the growth rate decreases monotonically as $r$ is decreased. It is evident from fig. \ref{fig8}(a) that zone $III$ is observed for relatively high density contrast fluid pairs, thus the decrement in density contrast with decrease in $r$ dominates over the decrement in surface tension. Fig. \ref{fig8}(b) is the culmination of the aforementioned growth rate behaviour in form of a regime-map demarcating the boundaries between different zones as a function of $\textit{A}$ and $log(\Web)$.
\\
The results obtained from the linear stability analysis are however limited to inviscid fluids. Thus, we opted for numerical route of investigations to first corroborate our findings and subsequently investigate late-time dynamics of viscous binary fluids elucidated in ensuing sections.

\subsubsection {Numerical solutions with negligible viscosity}

\begin{figure}
\centering 
\includegraphics[width=1.0\textwidth]{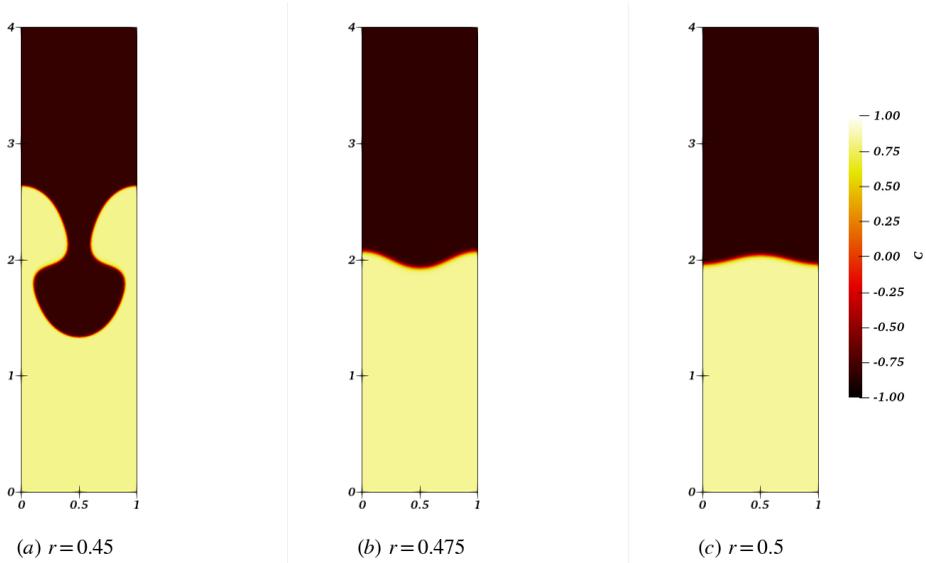} 
\caption{Determination of threshold $r$ for a fluid pair with $\textit{A} = 0.1$ and $\Web = 100$ from the interface topology at $t = 20$ for (a) $r = 0.45$, (b) $r = 0.475$ and (c) $r = 0.5$. The rest of the pertinent parameters are $\Rey = 10^{5}$, $h_{0} = 0.06$.}
\label{fig9}
\end{figure}

\begin{figure}
\centering 
\includegraphics[width=1.0\textwidth]{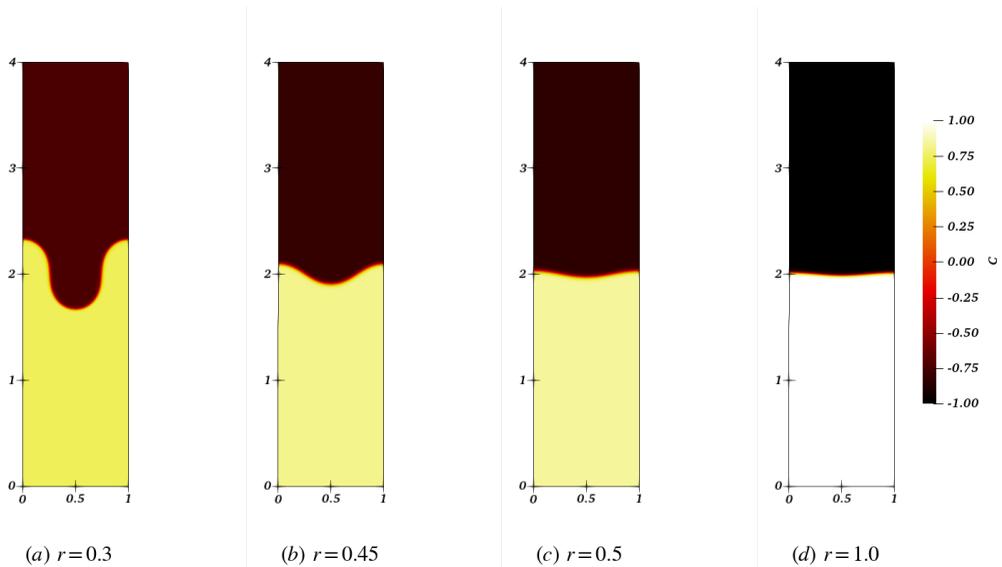} 
\caption{The interface topology for a fluid pair with $\textit{A} = 0.1$ and $\Web = 100$ at $t = 8$ for (a) $r = 0.3$, (b) $r = 0.45$, (c) $r = 0.5$ and (d) $r = 1.0$ . The rest of the pertinent parameters are $\Rey = 10^{5}$, $h_{0} = 0.06$.}
\label{fig10}
\end{figure}

\begin{figure}
\centering 
\includegraphics[width=0.5\textwidth]{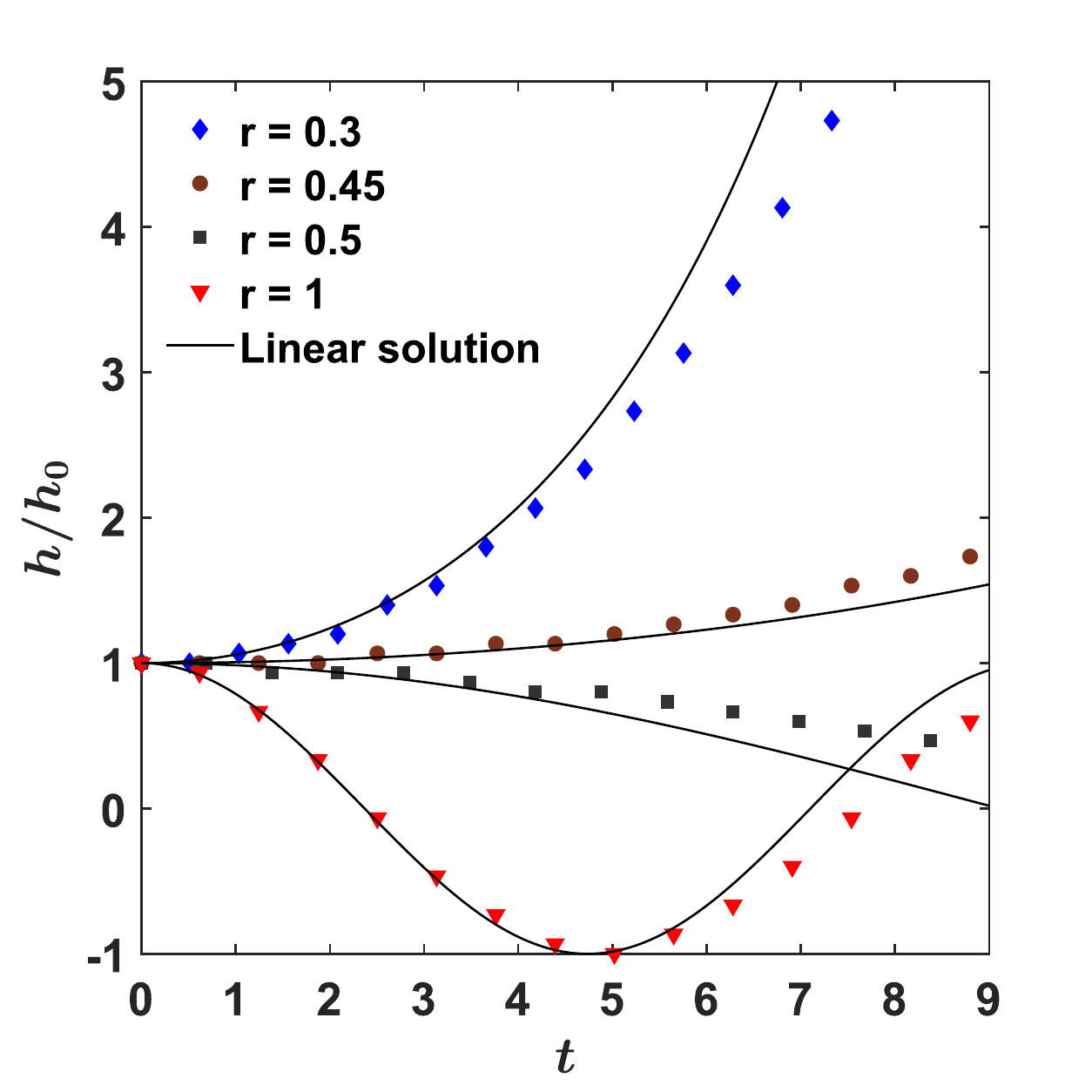} 
\caption{Comparison of numerical solution for perturbation amplitude growth with linear solution}
\label{fig11}
\end{figure}

The objective of this section is to numerically corroborate the findings of the inviscid linear stability analysis. The dimensions of the computational domain for this analysis is $[L \times 4L]$ as the linear theory holds during early-stages of perturbation growth \citep{Lewis1950}. The inviscid assumption is substantiated by considering $\Rey = 10^{5}$ in numerical simulations. The linearization of the equations is valid for small amplitude perturbations \citep{Chandrasekhar1961}. Thus, the interface is initialized with perturbation of amplitude $h_{0} = 0.06$ \citep{Celani2009}. A binary fluid pair in zone $I$, with properties $\textit{A} = 0.1$ and $\Web = 100$, is considered in the current analysis. Employing eq. \ref{cutoffMiscibCoeff_numericalWe}, the threshold value of the miscibility parameter $(r)$ predicted from the linear theory is $r_{th_{lin}} = 0.465$. Consequently, numerical simulations were performed to obtain the threshold $r$ as shown in fig. \ref{fig9} by investigating temporal evolution of the interface perturbation amplitude. It is evident that $r = 0.45$ leads to an unstable interface ascribed to the formation of spike and bubble whereas $r = 0.5$ leads to an oscillating interface thereby leading to gravity-capillary waves in case of infinitesimal viscosity. Notably, the configuration with $r = 0.475$ leads to steady-state solution with interface perturbation being independent of time. The destabilizing buoyancy force is exactly counteracted by the stabilizing surface tension force thereby emanating the state of marginal stability. The relative difference between the threshold values predicted from the linear stability analysis and the numerical simulations is $2.15\%$. This reasonably good agreement between the two methods of investigation establishes the validity and accuracy of our phase-field model.

Furthermore, a quantitative comparison of the growth rate $\alpha$ of interface perturbation in case of unstable configuration and the frequency of oscillation $\omega$ of the interface in case of stable configuration is performed. We continue our analysis with the same fluid pair with properties $\textit{A} = 0.1$ and $\Web = 100$. Four different values of  $r (= 0.3, 0.45, 0.5$ and $1.0)$ are considered. Fig. \ref{fig10} depicts the interface topology at $t = 8$. As expected, $r = 0.3$ and $0.45$ correspond to an unstable interface and $r = 0.5$ and $1.0$ correspond to a stable interface. Utilizing the potential theory, the amplitude of the perturbation for RT instability and gravity-capillary waves can be determined as follows \citep{Sharp1984}:

\begin{equation}
h(t) = h_{0}\cosh(\alpha t), 
\label{perturbationAmplitude_RTI}
\end{equation}

\begin{equation}
h(t) = h_{0}\cos(\omega t), 
\label{perturbationAmplitude_GCW}
\end{equation}

The solution obtained from eq. \ref{perturbationAmplitude_RTI} and \ref{perturbationAmplitude_GCW} were compared with the numerical solutions as shown in fig. \ref{fig11}. The amplitude perturbation was numerically calculated by tracking the iso-line of $c = 0$ at the centerline of the domain $x = 0.5$. The numerical solution agrees reasonably well with the linear solution during early-stages of the interface evolution. Notably, for $r = 0.3$, the perturbation growth slows down with time. This effect is attributed to the shear induced sharpening of the interface as the spike grows. A detailed discussion for the reasons behind this effect is provided in the next section.

\subsubsection {Numerical solutions with finite viscosity}

\begin{figure}
\centering 
\includegraphics[width=1.0\textwidth]{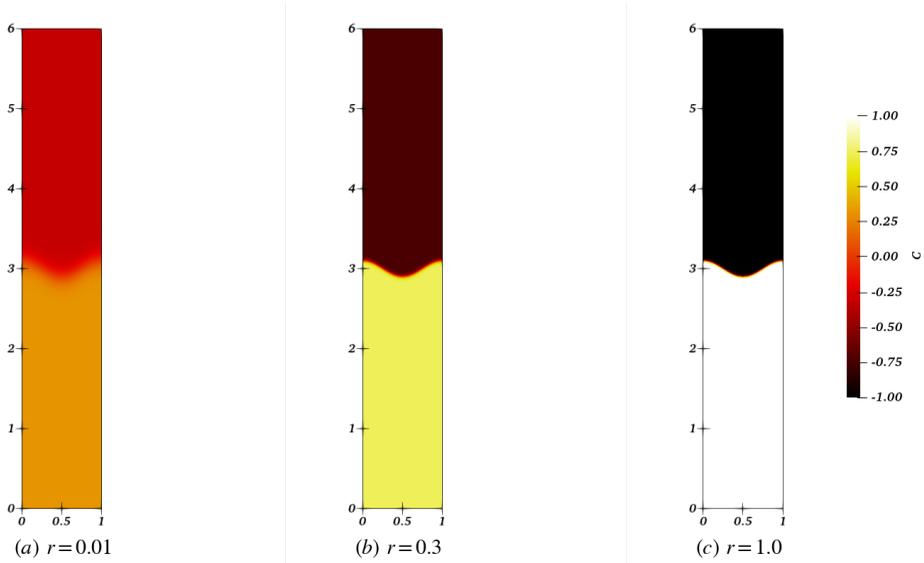} 
\caption{The initial interface topology for a fluid pair with $\textit{A} = 0.2$, $\Web = 1000$ and $\Rey = 5000$ with $h_{0} = 0.1$ for (a) $r = 0.01$, (b) $r = 0.3$ and (c) $r = 1.0$.}
\label{fig12}
\end{figure}

\begin{figure}
\centering 
\includegraphics[width=0.4\textwidth]{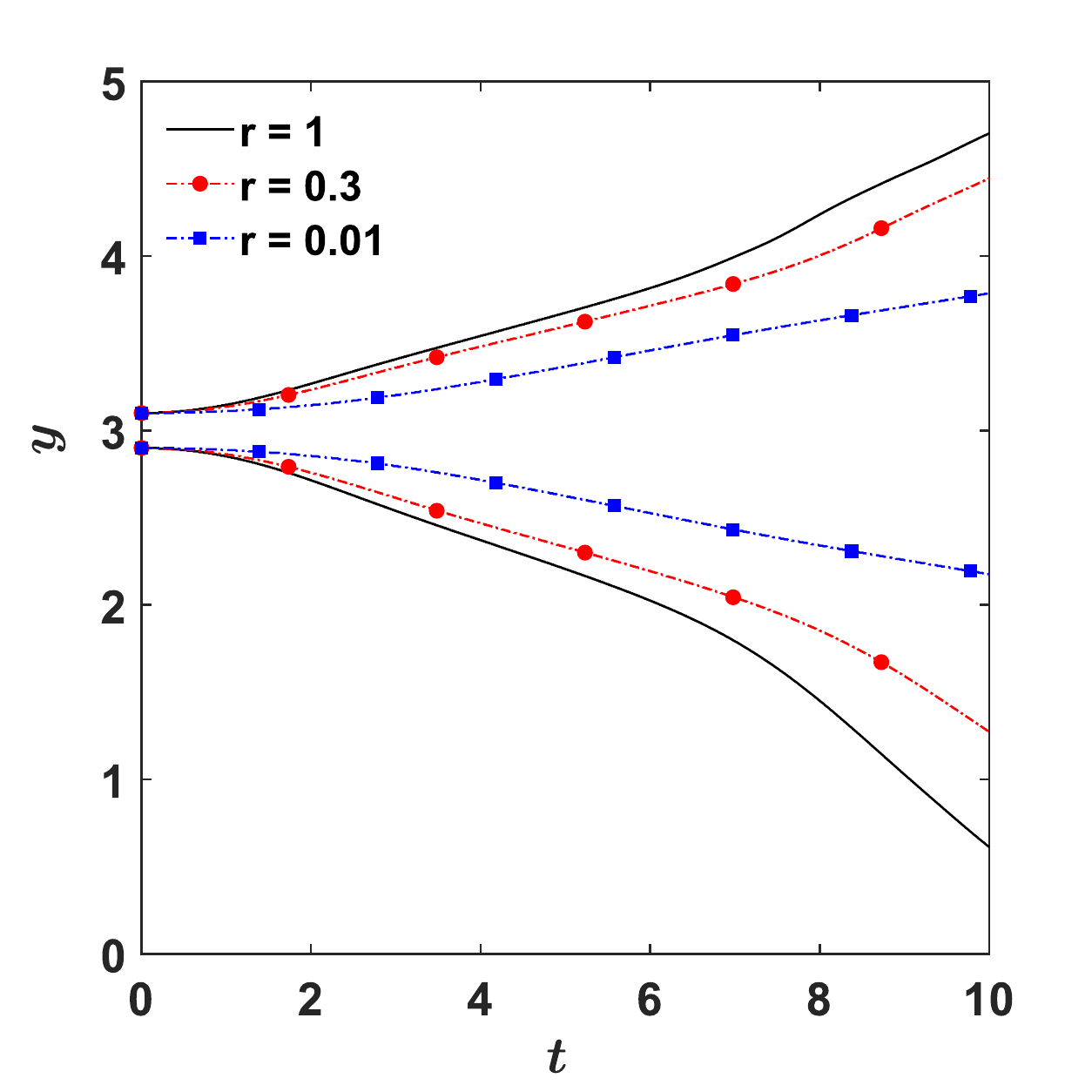} \hspace{0.2cm}
\includegraphics[width=0.4\textwidth]{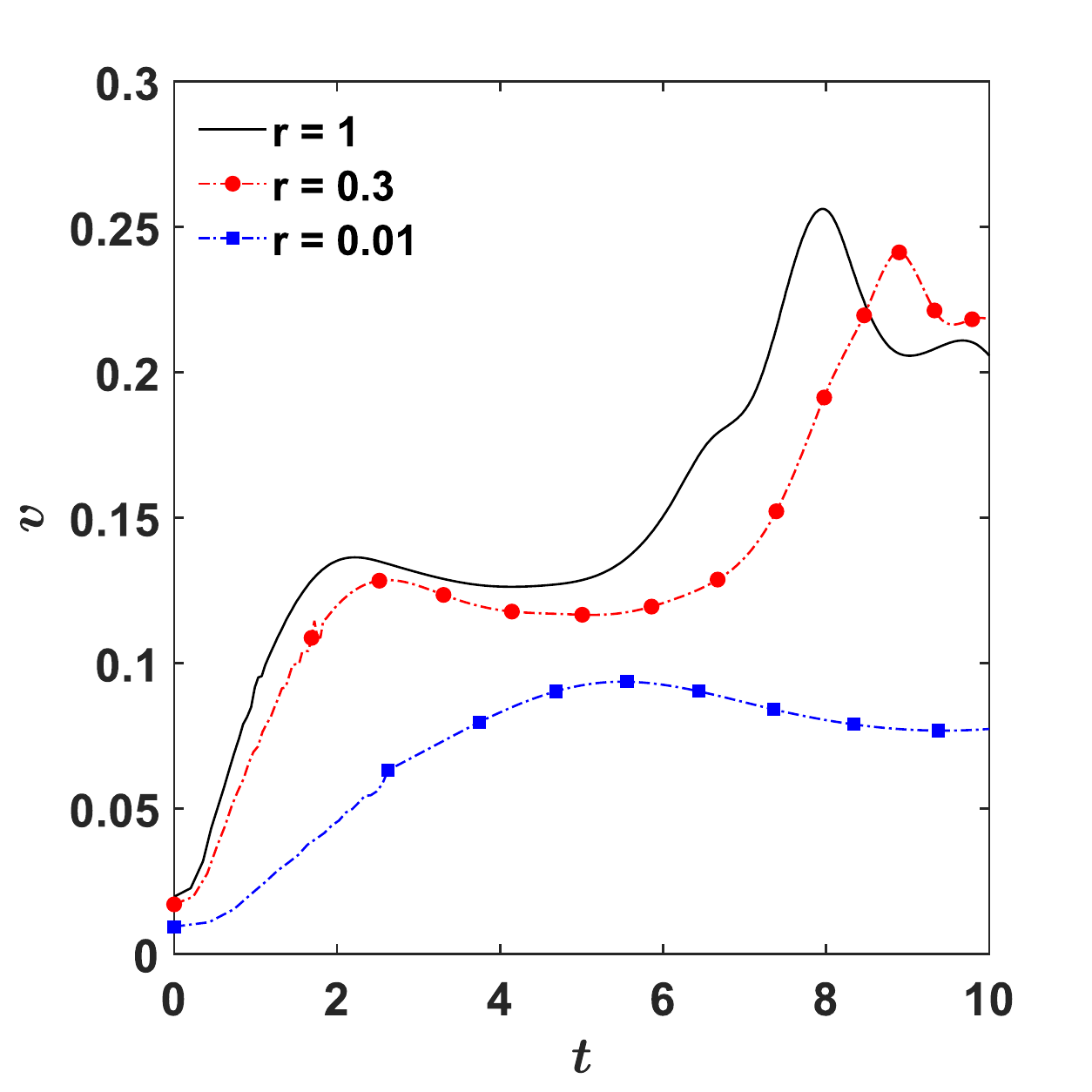} \\
\hspace{0.1cm} (a) \hspace{5.3cm} (b) \\
\caption{Temporal evolution of the (a) spike and bubble tip location and (b) bubble tip velocity for three different values of $r$ namely $ r = 0.01$, $r = 0.3$ and $r = 1.0$. The rest of the dimensionless parameters are $\textit{A} = 0.2$, $\Rey = 5000$ and $\Web = 1000$.}
\label{fig13}
\end{figure}

\begin{figure}
\centering 
\includegraphics[width=1.0\textwidth]{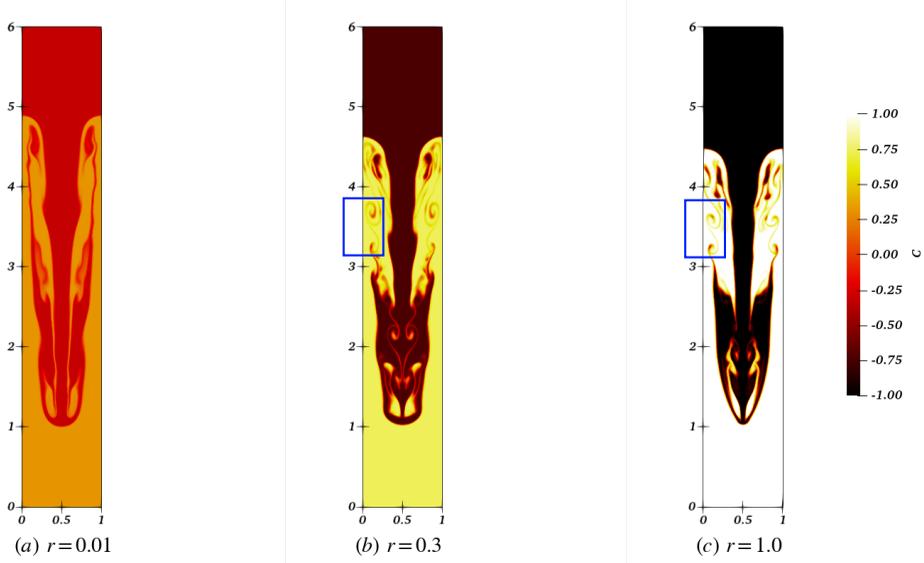} 
\caption{The interface topology for a fluid pair with $\textit{A} = 0.2$, $\Web = 1000$ and $\Rey = 5000$ with $h_{0} = 0.1$ at same spike location for (a) $r = 0.01$ at $t = 17.6$, (b) $r = 0.3$ at $t = 10.8$ and (c) $r = 1.0$ at $t = 9.0$.}
\label{fig14}
\end{figure}

\begin{figure}
\centering 
\includegraphics[width=1.0\textwidth]{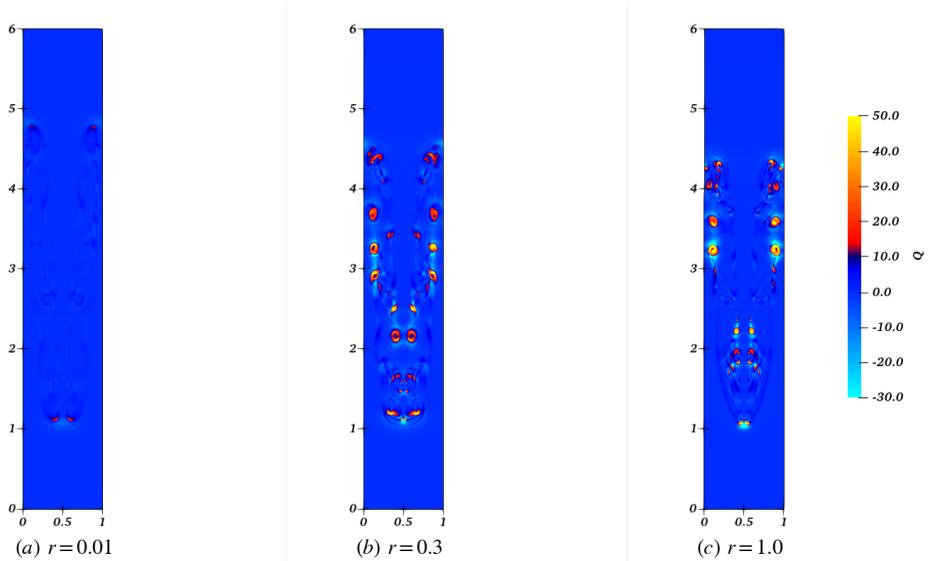} 
\caption{The Q-criteria for a fluid pair with $\textit{A} = 0.2$, $\Web = 1000$ and $\Rey = 5000$ with $h_{0} = 0.1$ at same spike location for (a) $r = 0.01$ at $t = 17.6$, (b) $r = 0.3$ at $t = 10.8$ and (c) $r = 1.0$ at $t = 9.0$.}
\label{fig15}
\end{figure}

Having developed the primary understanding of the RT instability in context of binary fluids, we now extend our investigations to relatively large-amplitude $(h_{0} = 0.1)$ perturbations on interface between fluids having finite viscosity. The viscosity ratio is maintained at unity $({\tilde{\mu}_{2} \over \tilde{\mu}_{1}} = 1)$ throughout the analysis. The RT instability in context of miscible fluids have traditionally been investigated with zero surface tension \citep{Ramprabhu2006}. Thus, the current study emphasizes on binary fluids exhibiting relatively higher surface tension at the immisicible limit. Furthermore, we analyze the late-time behavior of interface evolution as the proximity to the UCST is varied. To begin with, a binary fluid in zone $III$ with properties $\textit{A} = 0.2$, $\Web = 1000$, $\Rey = 5000$ at three different values of $r(= 0.01, 0.3$ and $1.0)$ is considered as shown in fig. \ref{fig12}. The divergence in the interfacial thickness is evident as the $r$ is decreased. The associated decrease in the concentration gradient causes a reduction in magnitude of capillary forces. The onset and the early-stage development of the RT morphology follows the behavior predicted from linear stability analysis as described in fig. \ref{fig8}. The temporal evolution of the tip of the falling spike and the rising bubble, as shown in fig. \ref{fig13}(a), demonstrates the monotonic decrease in the rate of growth of the perturbation amplitude as $r$ is decreased. This behaviour is further evident from the velocities of the rising bubble as shown in fig. \ref{fig13}(b). Notably, for $r = 0.3$ and $r = 1.0$, the rising bubble was found to follow ``acceleration-deceleration" phase \citep{Hu2019}  whereas for $r = 0.01$, the bubble velocity saturates during intermediate time stage. The acceleration of the rising bubble is attributed to the emanation of secondary instability in form of KH rolls \citep{Ramprabhu2012}, which reduces the frictional drag and engenders a vertical jet of momentum to propel the bubble forward. The KH rolls are advected towards the bubble head with the flow, where they eventually dissipate due to viscosity leading to deceleration phase. In order to present a comparative behavior of the secondary instability, the interface topology and Q-criteria for the three cases at same spike location is presented through fig. \ref{fig14} and fig. \ref{fig15} respectively. The insets in fig. \ref{fig14}(b,c) highlight the formation of the KH rolls for $r = 0.3$ and $r = 1.0$. The temporal evolution of the interface topology in all the three cases leads to shear induced sharpening of the interface to follow mass conservation principle. This effect was found to be more pronounced for $r = 0.01$ as observed by comparing the interface thickness in fig. \ref{fig12}(a) and fig. \ref{fig14}(a). The KH instability emanates when the interface presumes a sharp-enough topology therefore delaying the acceleration phase of the rising bubble in $r = 0.01$ significantly. Fig. \ref{fig15} demonstrates the higher propensity of formation of vortex-dominated flow regions, characterized by $Q > 0$, in $r = 1.0$ and $r = 0.3$ as compared to $r = 0.01$ which underpins the provocation of the secondary instability.

\begin{figure}
\centering 
\includegraphics[width=1.0\textwidth]{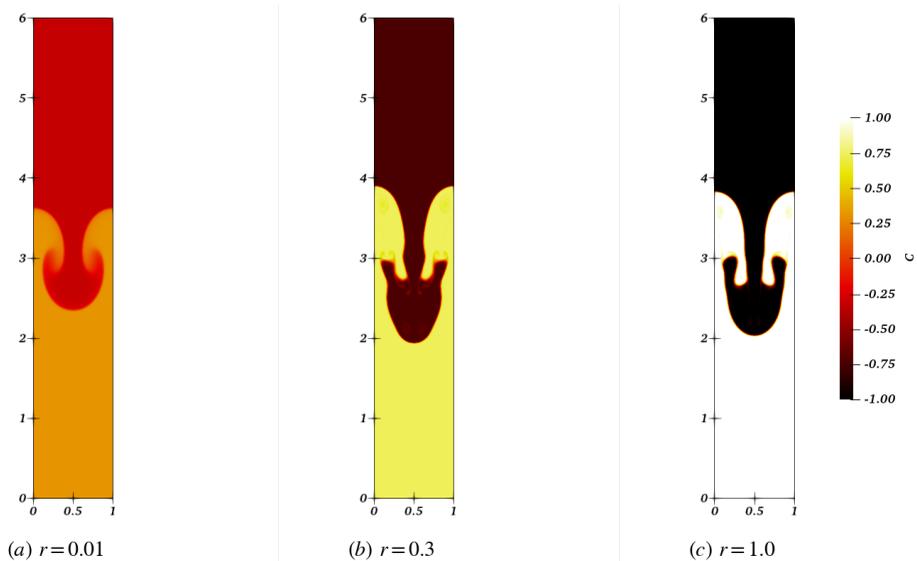} 
\caption{The interface topology for a fluid pair with $\textit{A} = 0.2$, $\Web = 100$ and $\Rey = 5000$ with $h_{0} = 0.1$ at $t = 8$ for (a) $r = 0.01$, (b) $r = 0.3$ and (c) $r = 1.0$.}
\label{fig16}
\end{figure}

\begin{figure}
\centering 
\includegraphics[width=1.0\textwidth]{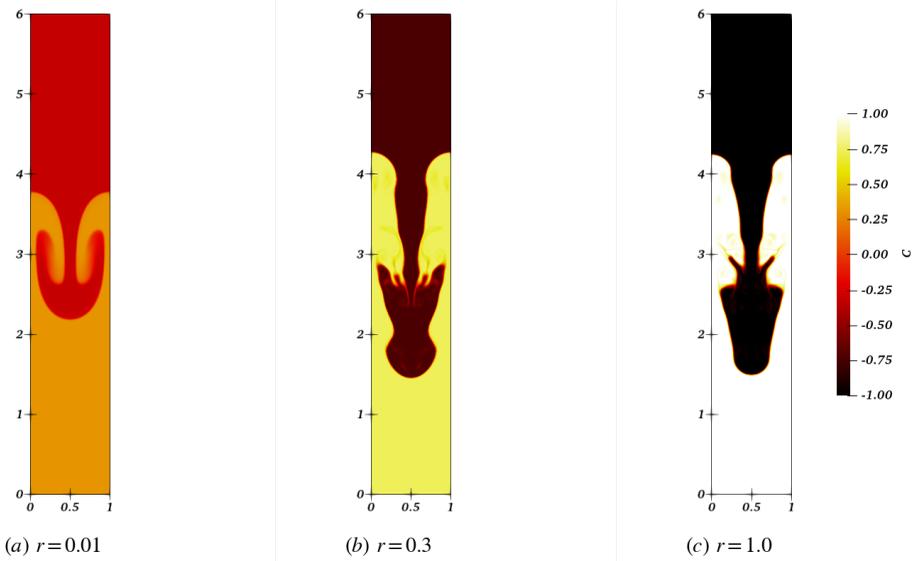} 
\caption{The interface topology for a fluid pair with $\textit{A} = 0.2$, $\Web = 100$ and $\Rey = 5000$ with $h_{0} = 0.1$ at $t = 10$ for (a) $r = 0.01$, (b) $r = 0.3$ and (c) $r = 1.0$.}
\label{fig17}
\end{figure}

\begin{figure}
\centering 
\includegraphics[width=0.4\textwidth]{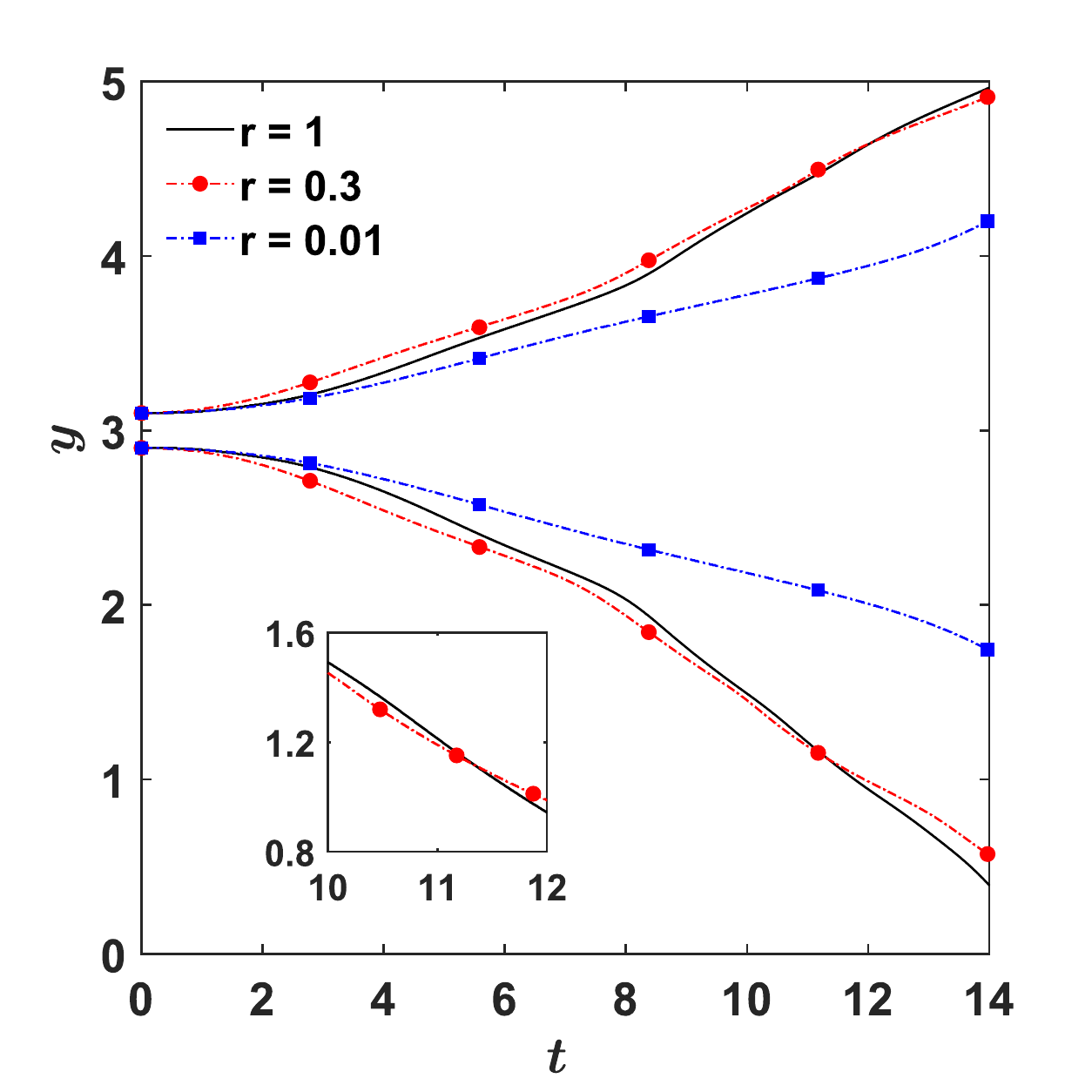} \hspace{0.2cm}
\includegraphics[width=0.4\textwidth]{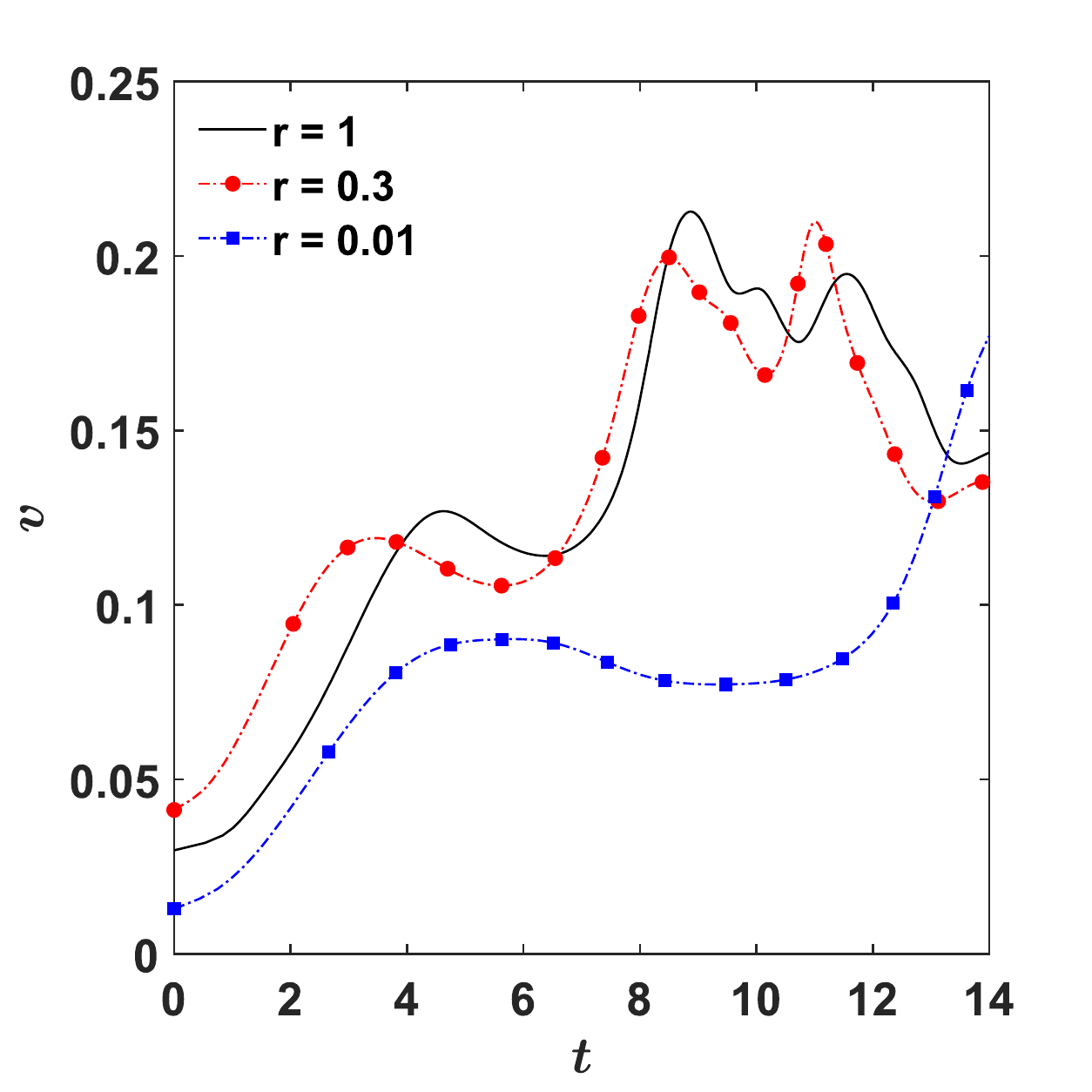} \\
\hspace{0.1cm} (a) \hspace{5.3cm} (b) \\
\caption{Temporal evolution of the (a) spike and bubble tip location and (b) bubble tip velocity for three different values of $r$ namely $ r = 0.01$, $r = 0.3$ and $r = 1.0$. The rest of the dimensionless parameters are $\textit{A} = 0.2$, $\Rey = 5000$ and $\Web = 100$.}
\label{fig18}
\end{figure}

\begin{figure}
\centering 
\includegraphics[width=1.0\textwidth]{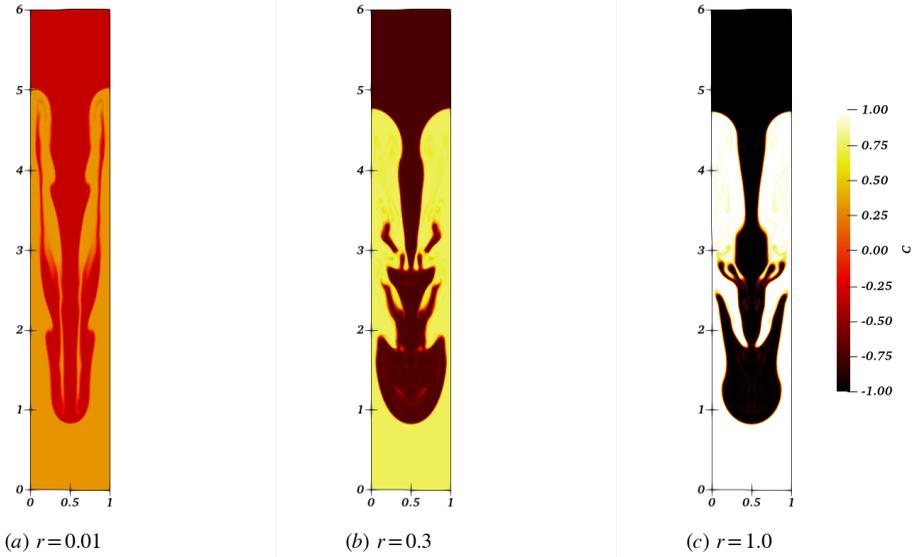} 
\caption{The interface topology for a fluid pair with $\textit{A} = 0.2$, $\Web = 100$ and $\Rey = 5000$ with $h_{0} = 0.1$ at same spike location for (a) $r = 0.01$ at $t = 18.3$, (b) $r = 0.3$ at $t = 12.9$ and (c) $r = 1.0$ at $t = 12.5$.}
\label{fig19}
\end{figure}

Finally, a binary fluid pair in zone $II$ was considered with aforementioned values of $r$ to complete the analysis in configuration $(\mathrm{1})$. The pertinent thermophysical properties are given by $\textit{A} = 0.2$, $\Web = 100$, $\Rey = 5000$. The numerical simulations corroborate the growth rate behaviour predicted from the linear stability analysis $(\alpha_{r_{0.3}} > \alpha_{r_{1.0}} > \alpha_{r_{0.01}})$ during early-stages of the interface evolution as demonstrated in fig. \ref{fig16} - \ref{fig18}. A novel regime during the temporal evolution of the interface topology for the falling spike is observed for $r = 1.0$ and $r = 0.3$ between $t = 8$ to $t = 10$. The formation of a vortex dominated region near the tip of the falling spike results in roll-up of the interface (see fig. \ref{fig16}). However, the relatively larger surface tension retracts the roll-up section to form a spike with single column with a protruding finger as shown in fig. \ref{fig17}(b,c). Fig. \ref{fig18}(a) depicts the location of the tip of the falling spike and the rising bubble. While the early-stage growth rate follows the qualitative trend expected from the zone $II$ binary fluids, the late-time dynamics offers contradictory behaviour for $r = 1$ and $r = 0.3$. The spike falls to the same distance at $t \approx 11.3$, as shown in the inset, with $r = 1$ falling faster than $r = 0.3$ subsequently. The velocity of the rising bubble (see fig. \ref{fig18}(b)) follows the same trend with early stage velocity being maximum for $r = 0.3$ till $t \approx 8.5$ except for a brief period of accelerating regime for $r = 1$ from $t = 3.9$ to $t = 6.5$. Post $t = 8.5$, the average velocity of the fall of spike for $r = 1$ is higher than $r = 0.3$. This anomaly in the behaviour can be explained through fig. \ref{fig19}. During the late-time stages of the RT instability in case of $r = 0.3$, the lighter fluid gets trapped in the spike of the heavier fluid unlike $r = 1$ as shown in fig. \ref{fig19}(b,c). This entrapment leads to buoyant force acting in the upward direction thereby reducing the net downward acceleration which in-turn slows down the falling spike. Consequently, the velocity of the rising bubble is reduced to ensure the mass conservation. The growth behaviour demonstrates week dependence on $\Web$ for $r = 0.01$. The extreme proximity to the UCST results in an almost homogenization of the concentration field. Thus, the capillary forces are very weak and no retraction of the spike roll-up is observed. The interface continues to evolve similar to the previous case (see fig. \ref{fig14}(a)). While on one hand, the homogenization reduces the density disparity thereby reducing early-stage growth rate, the reduction in surface tension on the other hand leads to the emanation of dominant secondary instability in form of KH rolls. Thus, the reacceleration phase is delayed yet prominent, eventually allowing the faster fall of spike in $r = 0.01$ at very-late time stages. Interestingly, a break-away segment of the falling spike is only observed in $r = 0.3$ as shown in fig. \ref{fig19}(b).\\

\subsection {Initialization with thermodynamic non-equilibrium}

\begin{figure}
\centering 
\includegraphics[width=1.0\textwidth]{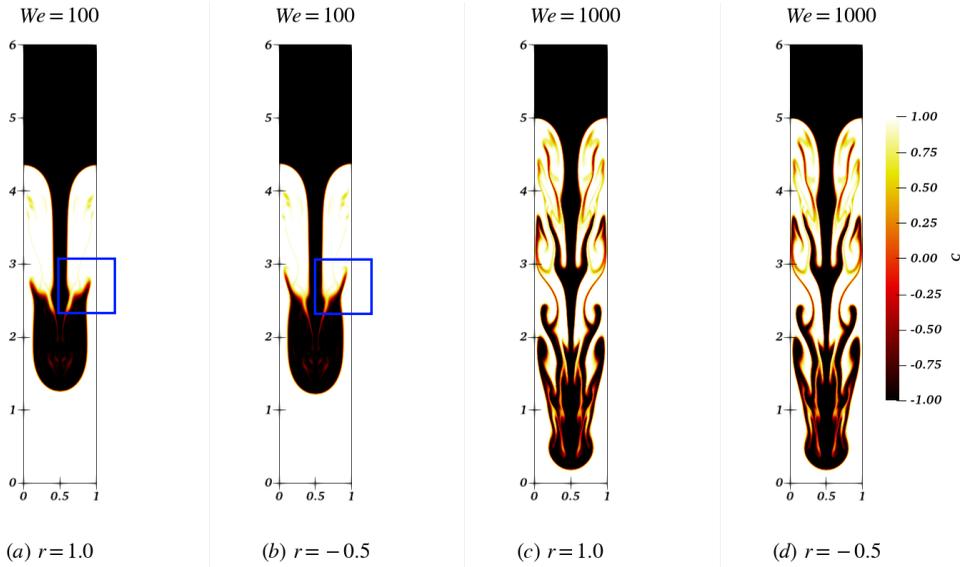} 
\caption{The interface topology for a fluid pair with $\textit{A} = 0.2$ and $\Rey = 1000$ with $h_{0} = 0.1$ at $t = 12$ for (a) $r = 1.0, We = 100$, (b) $r = -0.5, We = 100$, (c) $r = 1.0, We = 1000$ and (d) $r = -0.5, We = 1000$.}
\label{fig20}
\end{figure}

\begin{figure}
\centering 
\includegraphics[width=0.5\textwidth]{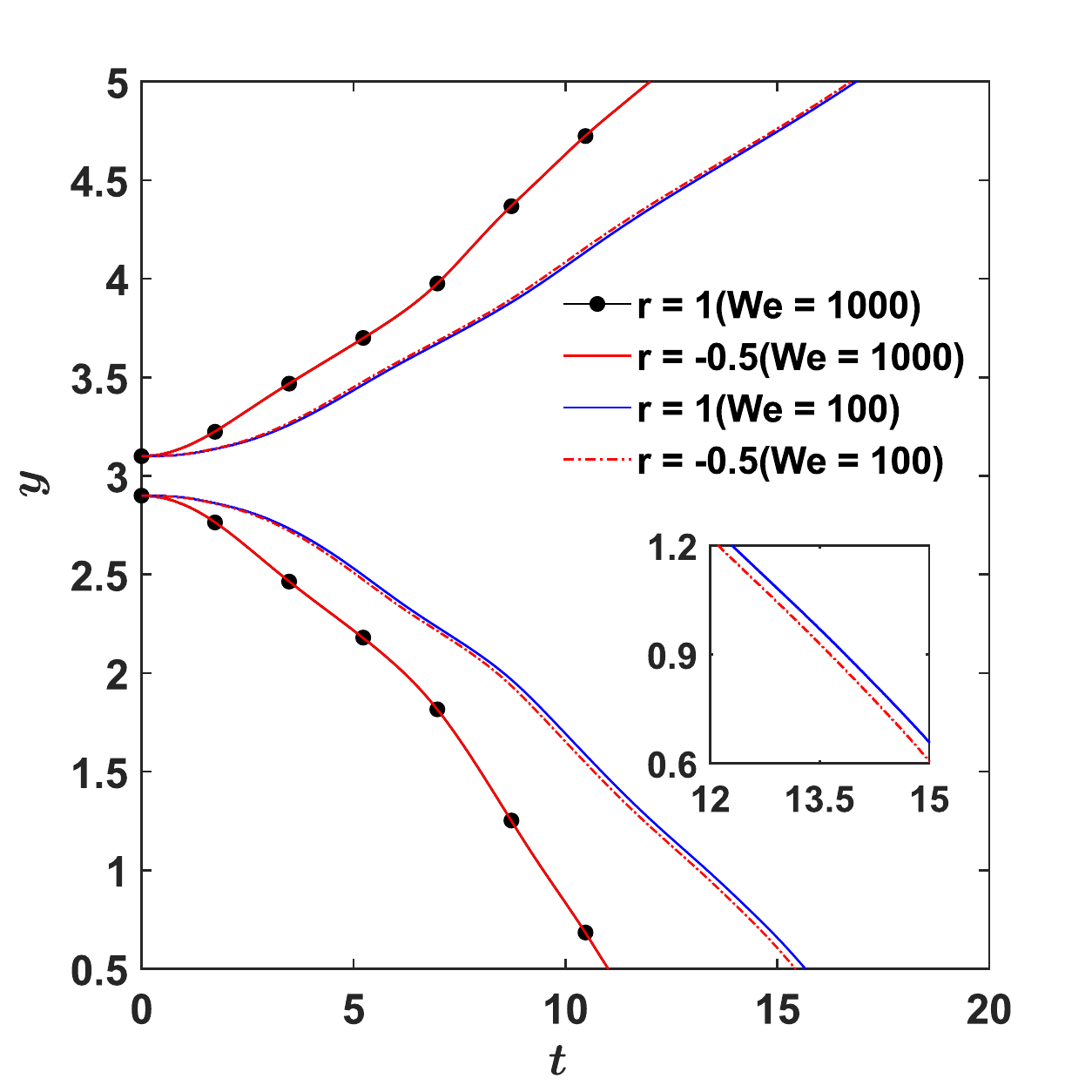} 
\caption{Temporal evolution of the spike and bubble tip location for 4 combination as follows:  $r = 1.0, We = 100$, $r = -0.5, We = 100$,  $r = 1.0, We = 1000$ and $r = -0.5, We = 1000$. The rest of the pertinent parameters are  $\textit{A} = 0.2$, $\Rey = 1000$ and $h_{0} = 0.1$ }
\label{fig21}
\end{figure}

In this section, we present our results for the configuration $(\mathrm{2})$ characterized by initialization of the system at a state of thermodynamic non-equilibrium therefore resulting in simultaneous advection as well as bilateral interfacial diffusion. A similar analysis performed by \citet{Lyubimova2019} was limited to confined domain. The objective of the current analysis is to investigate the effect of surface tension $\tilde{\sigma}_{0}$ on the development of the interface profile as the system is driven towards the state of thermodynamic equilibrium due to gradient of chemical potential. The two fluids are assumed to be just brought into contact therefore allowing no diffusion. The interface is allowed to evolve under two conditions - system temperature being far lower than the UCST $(r = 1)$ and system temperature being higher than UCST $(r = -0.5)$. It must be ensured that the system temperature is lower than the boiling point of the either liquid. The thermophysical properties of the fluids considered are given by $\textit{A} = 0.2$ and $\Rey = 1000$. Two different values of Weber number $\Web = 100$ and  $1000$ are considered to present a comparative analysis. The transformation of the system from the state of immiscibility $(r = 1)$ to miscibility $(r = -0.5)$ is characterized by the continuous evolution of the bulk free energy from the double-well potential function to single-well potential function (see fig. \ref{fig1}). However, the evolution is limited by the diffusion time scale. Consequently, the interface evolves in an identical manner irrespective of the system temperature for relatively low surface tension fluids as shown in fig. \ref{fig20}(c,d). Thus, our results corroborate the conclusions of \citet{Lyubimova2019} for $\Web = 1000$. Interestingly, for high surface tension fluids, $\Web = 100$, minute differences in the interface topology and the location of the falling spike is observed as evident from fig. \ref{fig20}(a,b). These differences are attributed to the high gradient in the chemical potential as it is a function of mixing energy density. The instantaneous location of the tip of the falling spike and the bubble for the four afore-mentioned configurations is presented in fig. \ref{fig21}. The inset in the fig. \ref{fig21} highlights the faster rate of fall of the spike for $r = -0.5$ as compared to $r = 1$ for $\Web = 100$ due to the advection aided diffusion.

\section{Conclusion}
\label{sec:conclusion}

In this study, we quantify the effect of system temperature on the isothermal RT instability in context of binary fluids exhibiting temperature sensitive miscibility. A novel phase field model, improving the shortcomings of our previous model (\citet{Bestehorn2021}; \citet{Borcia2022}) is proposed. The model utilizes a dimensionless parameter $r$ to quantify the system temperature relative to the upper critical solution temperature (UCST). The bulk free energy, a function of the miscibility parameter $r$, transforms from the double well potential to a single well potential as system temperature is raised above the UCST. The RT instability is investigated in two configurations namely, initialization at a thermodynamically stable state and initialization at thermodynamically unstable state. While the former configuration allows the system to attain thermodynamic equilibrium based on the system temperature ahead of the hydrodynamic instability, the later configuration assumes the fluids to be just brought into contact irrespective of system temperature therefore leading to simultaneous advection and bilateral interfacial diffusion.   \\
For the first configuration, the proposed phase-field model coupled with Boussinesq - inviscid Navier Stokes equations is employed to derive the dispersion relation for the gravity-capillary waves and the RT instability as a function of the miscibility parameter $r$. Utilizing the dispersion relation for RT instability, we demarcated the boundary between stable and unstable fluid configurations as a function of the parameter $r$. Further the dispersion relation is utilized to conduct parametric investigation of the early-stage perturbation growth characteristics. Subsequently, the Boussinesq approximation was relaxed to perform the numerical simulations. The numerical simulations corroborate the findings from the linear stability analysis to establish the validity and accuracy of our model. Three distinct zones of early-stage perturbation growth were identified having qualitatively different behaviour. The emanation of the secondary instability in form of KH rolls was observed. We found the strength of the KH rolls to be a strong function of the miscibility parameter $r$. The numerical investigation of the second configuration corroborated the results of \citet{Lyubimova2019} for low surface tension fluids.\\
These results are very encouraging, given the growing body of literature on binary fluids. The proposed phase-field model can be directly employed to tackle plethora of problems of practical interest. The observations from the current study, naturally motivates us to investigate the KH instability in the context of binary fluids, which shall form the subject matter for our subsequent study.

\backsection[Acknowledgements]{The authors gratefully acknowledge the computing time provided to them on the high-performance computer Lichtenberg II. This is funded by the German Federal Ministry of Education and Research (BMBF) and the State of Hesse.}

\backsection[Funding]{A.D. and S.A. gratefully acknowledge the support provided by CNES (Centre national d'études spatiales: grant 9384- 4500082826). C.H. and H.M. acknowledge the funding by Deutsche
Forschungsgemeinschaft (DFG, German Research Foundation) project number
237267381 SFB/Transregio 150 sub-project B08.}

\backsection[Declaration of interests]{The authors report no conflict of interest.}

\backsection[Data availability statement]{The datasets generated supporting the findings of this article
are obtainable from the corresponding author upon reasonable request.}

\backsection[Author ORCIDs]{\textbf{A. D.}, \href{https://orcid.org/0009-0007-1220-3357}{https://orcid.org/0009-0007-1220-3357}; \textbf{C. H.}, \href{https://orcid.org/0000-0003-2021-720X}{https://orcid.org/0000-0003-2021-720X}; \textbf{H. M.}, \href{https://orcid.org/0000-0001-8684-0681}{https://orcid.org/0000-0001-8684-0681}; \textbf{S. A.}, \href{https://orcid.org/0000-0002-7946-6350}{https://orcid.org/0000-0002-7946-6350}}

\backsection[Author contributions]{\textbf{A.D.} and \textbf{S.A.} conceptualized the problem. \textbf{A.D.} proposed the modified phase-field method, derived the dispersion relation, implemented the novel phase-field model, performed numerical simulations, analyzed the data, prepared the original manuscript, reviewed and edited the manuscript. \textbf{C.H.} helped \textbf{A.D.} to acquire the nuances of the original numerical solver, reviewed and edited the manuscript. \textbf{S.A.} and \textbf{H.M.} supervised the project, ensured funding and resources, reviewed and edited the manuscript.}

\appendix

\section{The equilibrium interface profile}\label{appA}

Rewriting the expression of chemical potential given by eq. \ref{chemicalPotential} 

\begin{equation}
\tilde{\phi} = {\delta \tilde{F} \over \delta c} = \tilde{f_{0}}^\prime(c) - \Lambda {\tilde{\nabla}^2}c, 
\label{rewritechemicalPotential}
\end{equation}

The state of thermodynamic equilibrium is characterized by zero chemical potential $(\phi = 0)$. The integration of the resultant equation for a one-dimensional planar interface gives:

\begin{equation}
\tilde{f_{0}}(c) = {\tilde{\Lambda} \over 2}\Bigl({\frac{dc}{d\tilde{y}}}\Bigr)^2 + Z, 
\label{EquilibriumCriteria}
\end{equation}

where $Z$ is the integration constant. Substituting the expression for $\tilde{f_{0}}(c)$ from eq. \ref{bulkFreeEnergy} and with adequate choice of the integration constant $Z (= {1 \over 4\tilde{\epsilon}^2}r^{(2-a)})$, the eq. \ref{EquilibriumCriteria} transforms to:

\begin{equation}
\frac{dc}{d\tilde{y}} = {1 \over \sqrt{2}(\tilde{\epsilon} / \sqrt{r})}\Bigl(r^{(a-1) \over 2}c^2 - r^{(1-a) \over 2}\Bigr), 
\label{IntermediateStep}
\end{equation}

which is further integrated in straightforward manner to give:

\begin{equation}
 c = - r^{(1-a) \over 2} \tanh\Bigl({{\tilde{y}- \tilde{y}_{0} \over {\sqrt{2} (\tilde{\epsilon} / \sqrt{r}) }}}\Bigr), 
 \label{reWriteEquilibInterfaceProfile}
\end{equation}

The surface tension, $\sigma$, is then calculated by re-writing eq. \ref{surfaceTensionTheory} for an interface under equilibrium as:  

\begin{equation}
 \tilde{\sigma} = \int_{-\infty}^{\infty} {\tilde{\Lambda}}\Bigl(\frac{dc}{d\tilde{y}}\Bigr)^2 d\tilde{y}, 
 \label{reWriteSurfaceTensionTheory}
\end{equation}

Employing the expression for order parameter, $c$ (eq. \ref{reWriteEquilibInterfaceProfile}), the eq. \ref{reWriteSurfaceTensionTheory} is integrated to obtain:

\begin{equation}
 \tilde{\sigma} = {{2 \sqrt{2}} \over 3} {\tilde{\Lambda} \over \tilde{\epsilon}} r^{(3-2a) \over 2}. 
 \label{reWriteSurfaceTensionExpression}
\end{equation}

\section{The dispersion relation}\label{appB}

\citet{Celani2009} presented a thorough derivation of the dispersion relation pertaining to gravity-capillary waves with phase-field approach in the context of immiscible fluids. Herein, we briefly recall the methodology and present the details of the deviation arising due to the consideration of binary fluids. 
Let us start our analysis from the inviscid momentum conservation equation (eq. \ref{BoussinesqMomentumEquation}):

\begin{eqnarray}
{\partial \textbf{u} \over \partial t}+ \textbf{u} \bcdot \bnabla  (\textbf{u}) & = & -\bnabla P - \textit{A} c \textbf{g}
 \nonumber\\
 && \mbox{}  + {{3 \over {2 \sqrt{2}}}{1 \over {\Web_{B}\Cahn}}} \Bigl({|r|^a}c^3 - rc - \Cahn^2 {\nabla^2}c\Bigr)\bnabla c, 
\label{reWriteBoussinesqMomentumEquation}
\end{eqnarray}

where $c$ is perturbed with small amplitude disturbance to take the profile (For mathematical simplicity, let us assume $y_{0} = 0$ in eq. \ref{perturbedInterfaceProfile}):

\begin{equation}
 c = - r^{(1-a) \over 2} \tanh\Bigl({{y - h(x,t) \over {\sqrt{2} {(\Cahn / \sqrt{r})}}}}\Bigr), \label{reWritePerturbedInterfaceProfile}
\end{equation}

with $h(x,t=0) = h_{0}cos(kx)$. This perturbation about the state of thermodynamic equilibrium renders the chemical potential to be:

\begin{equation}
\phi =  - \Cahn^2 {\frac{\partial^2 c}{\partial x^2}}, 
\label{perturbedchemicalPotential}
\end{equation}

Linearizing the momentum equation (eq. \ref{reWriteBoussinesqMomentumEquation}) gives:
\begin{equation}
{\partial \textbf{u} \over \partial t} = -\bnabla P - \textit{A}c \textbf{g} - {{3 \over {2 \sqrt{2}}}{1 \over {\Web_{B}\Cahn}}} \Bigl(\Cahn^2 {\frac{\partial^2 c}{\partial x^2}}\Bigr) \bnabla c , 
\label{linearizedBoussinesqMomentumEquation}
\end{equation}

With these expressions for the order parameter $c$ (eq. \ref{reWritePerturbedInterfaceProfile}) and the chemical potential
$\phi$ (eq. \ref{perturbedchemicalPotential}), the linearized momentum equation (eq. \ref{linearizedBoussinesqMomentumEquation}) is integrated in the vertical direction to give (see supplementary material):

\begin{equation}
{\partial_t  \int_{-\infty}^{\infty} {v}dy} =   \bigl( {1 \over \Web_{B}} r^{(3-2a) \over 2}\bigr)h^{\prime\prime} - 2\textit{A} {g} r^{(1-a) \over 2}h  , 
\label{derive1dispersionRelationGCW}
\end{equation}

The vertical growth of the perturbation needs to match the corresponding interfacial velocity $(v^{int})$, thus giving: 
\begin{equation}
{\partial_t h} =  v^{int}(x,t)  , 
\label{derive2dispersionRelationGCW}
\end{equation}

Finally, the interfacial velocity is related to the integral in eq. \ref{derive1dispersionRelationGCW} by employing the Fourier transform \citep{Celani2009} to give the dispersion relation as a function of the miscibility parameter $r$ as: 

\begin{equation}
\omega^2(k) = {k^3 \over 2\Web_{B}}r^{(3-2a) \over 2} + k\textit{A}gr^{(1-a) \over 2}, 
\label{reWritedispersionRelationGCW}
\end{equation}

which reduces to the classical dispersion relation \citep{Chandrasekhar1961} in the limit of immiscibility $(r \to 1)$. The same fluid pair can be configured in an unstable density stratification if the direction of the external acceleration is reversed. This transforms the above-mentioned dispersion relation (eq. \ref{reWritedispersionRelationGCW}) as follows: 

\begin{equation}
\omega^2(k) = {k^3 \over 2\Web_{B}}r^{(3-2a) \over 2} - k\textit{A}gr^{(1-a) \over 2}, 
\label{dispersionRelationRT}
\end{equation}

Further, this would lead to growth of the initial perturbation if the effect of external acceleration dominates over the interfacial tension thereby resulting in RT instability . The growth rate $\alpha$ of the perturbation, which follows the relation ($\omega = \iota \alpha$), is then given by: 

\begin{equation}
\alpha^2(k) = k\textit{A}gr^{(1-a) \over 2} - {k^3 \over 2\Web_{B}}r^{(3-2a) \over 2},
\label{App_dispersionRelationRTI}
\end{equation}

Eq. \ref{dispersionRelationRTI} gives the dispersion relation for the RT instability in the context of binary fluids as a function of the miscibility parameter $r$. \\ \\

\bibliographystyle{jfm}
\bibliography{main}
%Use of the above commands will create a bibliography using the .bib file. Shown below is a bibliography built from individual items.

%% End of file `jfm2esam.bib'.

\end{document}